\documentclass[prc,
floatfix]{revtex4-1}
\usepackage[utf8]{inputenc}
\usepackage{amsmath}
\usepackage{amsfonts}
\usepackage{amssymb}
\usepackage{graphicx}
\usepackage{fullpage}
\usepackage{mathtools}
\usepackage{bm}
\usepackage{subfig}

\DeclareMathOperator*{\argmax}{argmax} 
\DeclareMathOperator*{\argmin}{argmin} 

\begin{document}
\title{Uncertainty quantification  of an empirical shell-model interaction using principal component analysis}

\author{Jordan M. R. Fox}
\email{jfox@sdsu.edu}
\author{Calvin W. Johnson}
\email{cjohnson@sdsu.edu}
\author{Rodrigo Navarro Perez}
\email{rnavarroperez@sdsu.edu}
\affiliation{%
 San Diego State University, San Diego, California, USA
}
\date{\today}
%

\begin{abstract}
    Recent investigations have emphasized the importance of uncertainty quantification (UQ) 
    in nuclear theory. We carry out UQ for configuration-interaction shell model calculations in the $1s$-$0d$ valence space, investigating the sensitivity of observables to perturbations in the 66 parameters (matrix elements) of a high-quality empirical interaction.  The large parameter space makes computing the corresponding Hessian numerically costly, so we compare a cost-effective approximation, using the Feynman-Hellmann theorem, to the full Hessian and find it works well. Diagonalizing the Hessian yields the principal components of the interaction: linear combinations of parameters ordered by sensitivity. This approximately decoupled distribution of parameters facilitates theoretical 
    uncertainty propagation onto structure observables: electromagnetic transitions, Gamow-Teller decays, and dark matter-nucleus scattering matrix elements.
\end{abstract}

\maketitle



\section{Introduction}


Recent advancements in nuclear theory have emphasized the importance of theoretical uncertainty quantification (UQ) \cite{dobaczewski2014error} with applications to, among other things, 
the nuclear force and effective field theory \cite{wendt2014uncertainty,furnstahl2015recipe,carlsson2016uncertainty,perez2016uncertainty,PhysRevC.100.044001,wesolowski2019exploring}, the optical model \cite{PhysRevC.95.024611, NavarroPerez:2018qzd}, density functional theory \cite{schunck2015quantification,PhysRevLett.114.122501}
and  the configuration-interaction shell model \cite{PhysRevC.96.054316,PhysRevC.98.061301}.
%

The shell model, which provides a useful conceptual framework for nuclear 
structure, 
can be approximately divided into \textit{ab initio} and empirical/phenomenological 
approaches.  \textit{Ab initio} calculations, such as the no-core shell model \cite{navratil2000large,barrett2013ab}, typically use forces built upon chiral effective field theory \cite{PhysRevC.49.2932} and thus are arguably more fundamental and also have been subject to considerable 
UQ \cite{wendt2014uncertainty,furnstahl2015recipe,carlsson2016uncertainty,perez2016uncertainty,PhysRevC.100.044001,wesolowski2019exploring}, but are limited to light nuclei,
approximately mass number $A<16$.  
Empirical shell model calculations \cite{BG77,br88,ca05} have a long, rich, and successful history, and, importantly, have been applied to a wide range of 
nuclei far beyond the $0p$ shell, but the theoretical underpinnings are more heuristic: 
individual interaction matrix elements in the lab frame (single-particle coordinates)
are adjusted to reproduce experimental data.

(We will  not consider here related but distinct 
methodologies such as coupled clusters \cite{hagen2010ab}, and we note but do not comment further on efforts to construct 
interactions that `look like' traditional empirical calculations but are derived with significant 
rigor from \textit{ab initio} forces \cite{doi:10.1146/annurev-nucl-101917-021120}.)


Previous work on UQ in the shell model focused on $0p$-shell calculation: one considered a 
simple interaction with only seven parameters, examining correlations using 
a singular-value-decomposition analysis \cite{PhysRevC.96.054316}; while the other used 
 17 parameters but did not consider correlations between parameters \cite{PhysRevC.98.061301}.

Because of the broad applications and demonstrated utility of the empirical shell model,  we carry out a sensitivity analysis on an widely-used, `gold standard' empirical shell-model interaction, Brown and Richter's universal $sd$-shell interaction, version B, or 
USDB \cite{PhysRevC.74.034315}.
Here, `$sd$-shell' means the valence space is limited to $1s$ and $0d$ single-particle orbits, 
with an inert $^{16}$O core.

In fitting their interaction, Brown and Richter followed a standard procedure \cite{BG77}. They minimized the total error with respect to experiment, defined as the $\chi^2$-function  in Eq.~(\ref{chireduced}) below, by taking the first derivatives with 
respect to the parameters, which yield the linear response of calculated energies to perturbations 
of the parameters, and then carried out gradient descent on the  independent 
parameters, here 63 two-body matrix elements and three single-particle energies. In the fit they found that about five or six linear combinations of parameters,
found by singular value decomposition as we do below,
were the most important. (Interesting, a similar result was found for random 
values of the matrix elements \cite{PhysRevC.81.054303}). 
Brown and Richter actually produced two interactions \cite{PhysRevC.74.034315}, USDA, which 
was found by fitting the first 30 linear combinations from singular value decomposition, 
and USDB, found by fitting 56 linear combinations.

For a Bayesian sensitivity analysis, discussed more fully in Appendix A, one must characterize the likelihood function for model parameters. In Laplace's approximation, one assumes the likelihood is well approximated by a Gaussian, which corresponds to a quadratic 
expansion in the $\chi^2$-function.
Even so, the matrix of second derivatives of $\chi^2$ (which, more rigorously, is 
the log-likelihood), or the \textit{Hessian},  needed is quite demanding to obtain.

We therefore consider a further simplification, approximating the Hessian by the same linear response (first derivatives of the energies), which are efficiently computed by the Feynman-Hellmann theorem \cite{hellman1937einfuhrung,PhysRev.56.340}.  As discussed below, this principal component analysis of the sensitivity is, in this approximation, singular value decomposition of the linear response. Importantly, we find that numerical corrections to the linear response matrix are small, making this approximation appealing for studying larger spaces wherein the full numerical calculation is too costly.

\section{The empirical configuration-interaction shell model}

We formally represent the nuclear Hamiltonian in second quantization, 
with $r,s,t,u$ labeling single-particle states, 
\begin{equation}
    \label{ham}
    \hat{{\cal H}} = \sum_{rs} T_{rs} \hat{a}^\dagger_r \hat{a}_s + \frac{1}{4}
    \sum_{rstu} V_{rs,tu}\hat{a}^\dagger_r \hat{a}^\dagger_s \hat{a}_u \hat{a}_t ,
\end{equation}
where typically one takes $T_{rs}$ as diagonal \textit{single-particle energies}, and 
the $V_{rs,tu}$ are two-body matrix elements.   As input to nuclear configuration-interaction codes, the two-body matrix 
elements are always coupled up to an angular momentum scalar so 
that the many-body angular momentum $J$ is a good quantum number of eigenstates \cite{BG77}.
(To be specific, the two-body matrix elements are $V_{JT}(ab,cd) = \langle ab; JT| \hat{V} | cd; JT\rangle$, where $\hat{V}$ is the nuclear two-body interaction and $| ab; JT \rangle$ 
is a normalized two-body state with nucleons in single-particle orbits labeled by $a,b$ coupled
up to total angular momentum $J$ and total isospin $T$.) In this paper, the single-particle 
energies and the coupled two-body matrix elements are the input parameters.

With the Hamiltonian (\ref{ham}) we want to find specific eigenpairs
\begin{equation}
    \hat{\cal H} | \psi_\alpha \rangle = E_\alpha | \psi_\alpha \rangle,
    \label{schrodinger}
\end{equation}
in this case low-lying states with experimentally known energies.  This is done by the 
configuration-interaction (CI) many-body method, which expands the wave function in a basis $\{ | a \rangle \}$, usually 
orthonormal, 
\begin{equation}
    | \psi_\alpha \rangle = \sum_a c_{\alpha, a} | a \rangle.
    \label{CIexpansion}
\end{equation}
Here $\alpha$ labels the eigenstates and their observables, in particular the energy $E_\alpha$.
For the basis we use the occupation representation of Slater determinants, that is, antisymmetrized products of single-particle states. We furthermore use basis states with fixed total $J_z$, also called an $M$-scheme basis.  By computing the matrix elements of the Hamiltonian in this same basis, 
${\cal H} _{a,b}= \langle a |   \hat{{\cal H}} | b \rangle,$ the Schr\"odinger equation 
(\ref{schrodinger}) is now a matrix eigenvalue problem, which we solve by the standard 
 Lanczos algorithm \cite{Lanczos} to 
extract the extremal eigenpairs of interest,
See \cite{BG77,br88,ca05} for a multitude of important and interesting details, 
and \cite{BIGSTICK,johnson2018bigstick} for information on the code used.

We assume a frozen $^{16}$O core and use the $1s$-$0d$ single-particle valence space, also called
the $sd$-shell. Assuming 
both angular momentum $J$ and isospin $T$ are good quantum numbers, one has only three independent 
single-particle energies and 63 independent two-body matrix element, for a total of 66 parameters.
Because each of those parameters appears linearly in the Hamiltonian, we can write
\begin{equation}
    \hat{\cal H} = \sum_{i} \lambda_i \hat{\cal O}_i
    \label{Hamiltonian_as_sum}
\end{equation}
where $\hat{\cal O}_i$ is some dimensionless one- or two-body operator.  Thus the parameters 
$\bm{\lambda}$ have dimensions of energy.

The set of parameters $\bm{\lambda} = \{\lambda_i\}$ we use are Brown and Richter's universal $sd$-shell interaction version $B$ (USDB) \cite{PhysRevC.74.034315}, which, along with its sister interaction USDA, are the current ``gold standards'' for empirical $sd$-shell calculations. The present study seeks to extend this model by computing theoretical uncertainties on model parameters and shell-model observables \cite{PhysRevC.78.064302,PhysRevC.80.034301}. While the parameter vector $\bm{\lambda}$ is formally considered a random variable, note that all calculations are performed about the USDB values.

An important nuance in using the USDB parameters is that while the single-particle energies are fixed, the 
two-body matrix elements are scaled by $(A_0/A)^{0.3}$, where $A$ is the mass number of the nucleus, and $A_0$ is 
a reference value, here $=18$. We account for this  by 
modifying (\ref{Hamiltonian_as_sum}) as $\hat{\cal H} = \sum_{i} \lambda_i (A_0/A)^{0.3}\hat{\cal O}_i$ (but only for the two-body matrix elements), so that we implicitly varied the parameters fixed at $A=18$.

Experimental energies in this paper are the same used in the original fit of the USDB Hamiltonian: absolute energies, relative to the $^{16}$O core and with Coulomb differences subtracted, of 608 states in 77 nuclei with $A = 21$ - $40$. The data  excludes any experimental uncertainties greater than 200 keV, and most are smaller, on the order of 10 keV. 

In the rest of this paper, we estimate the uncertainty in the USDB parameters and, from those,
estimate uncertainties in observables such as energies,  probabilities for selected electromagnetic 
and weak transitions, and for a matrix element relevant to dark matter direct detection.

\section{Sensitivity analysis}

\label{sensitivity}

Our analysis can be cast in terms most physicists are familiar with, see e.g. \cite{lyons1989statistics,PhysRevD.50.1173,press1992numerical,dobaczewski2014error}. In the Appendix we discuss the relationship to 
Bayesian analysis, giving a more rigorous starting point and setting the stage for more sophisticated analyses.

We begin with the $\chi^2$ function of parameters $\bm{\lambda}$, which is the usual sum of squared residuals over $N$ data:
\begin{equation}
		\chi^2(\bm{\lambda}) = \sum_{\alpha=1}^{N} \left(  \frac{E_\alpha^{SM}(\bm{\lambda}) - E_\alpha^{exp}}{\Delta E_\alpha} \right)^2,
		\label{chireduced}
\end{equation}
In addition $ E^{exp}_\alpha $ is the experimental excitation energy given in the data set and $ E^{SM}_\alpha(\bm{\lambda}) $ is the shell model calculation for that energy using the parameters $ \bm{\lambda} $. The total uncertainty $\Delta E_\alpha $  on the residual is expressed as experimental uncertainty $\Delta E_\alpha^{exp}$ and some \textit{a priori} theoretical uncertainty $\Delta E^{th}$ added in quadrature:
\begin{equation}
    \Delta E_\alpha^2  = (\Delta E^{th}) ^2  + (\Delta E_\alpha^{exp}) ^2
    \label{add_errors}
\end{equation}
Here we introduce $\Delta E^{th}$ as an estimated 
uncertainty on the shell-model predictions of the data. We assume it is 
independent of the level, that is, of $\alpha$, and fix it by requiring the reduced sum of squared 
residuals  $\chi^2_{\nu} = \frac{1}{\nu} \chi^2 \approx 1$ \cite{PhysRevD.50.1173}, which gives us $\Delta E^{th} \approx 150$ keV. Here $\nu$ is the number of degrees of freedom: the number of data points minus the number of parameters.
In their original paper, Brown and Richter set $\sigma^\mathrm{th}$ (equivalent to our 
$\Delta E^{th}$) $=0.1$ MeV as ``close to the rms value'' they eventually found, 126 keV 
\cite{PhysRevC.74.034315}.

\begin{figure}
  
    \includegraphics[scale=0.7,clip]{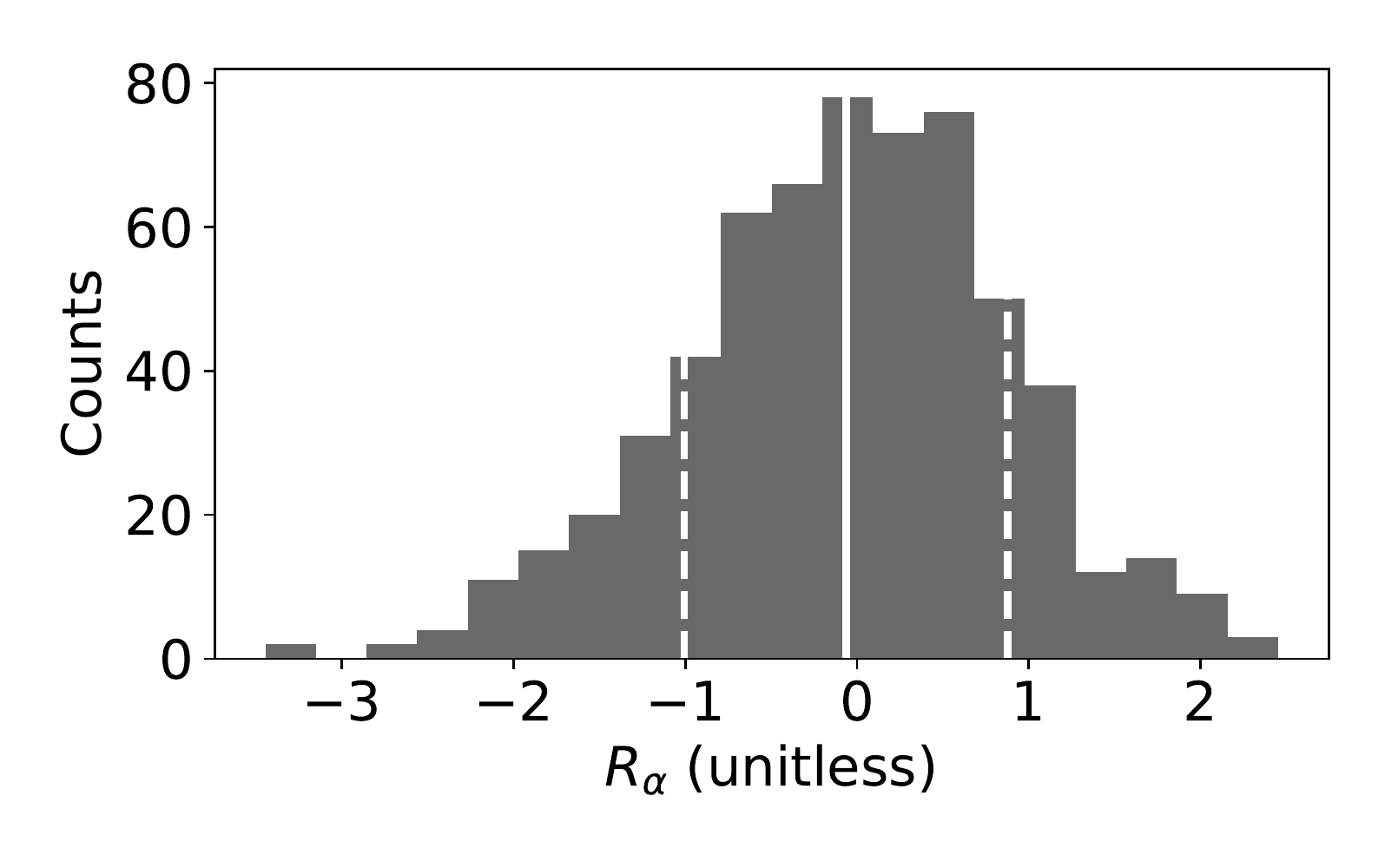}
    \label{fig:residual_hist}
\caption{Histogram of energy residuals $R_\alpha = (E_\alpha^{SM}(\bm{\lambda_\text{USDB}}) - E_\alpha^{exp})/\Delta E_\alpha$.}
\label{fig:residuals}
\end{figure}

\begin{figure}
    \centering
    \includegraphics[scale=0.7,clip]{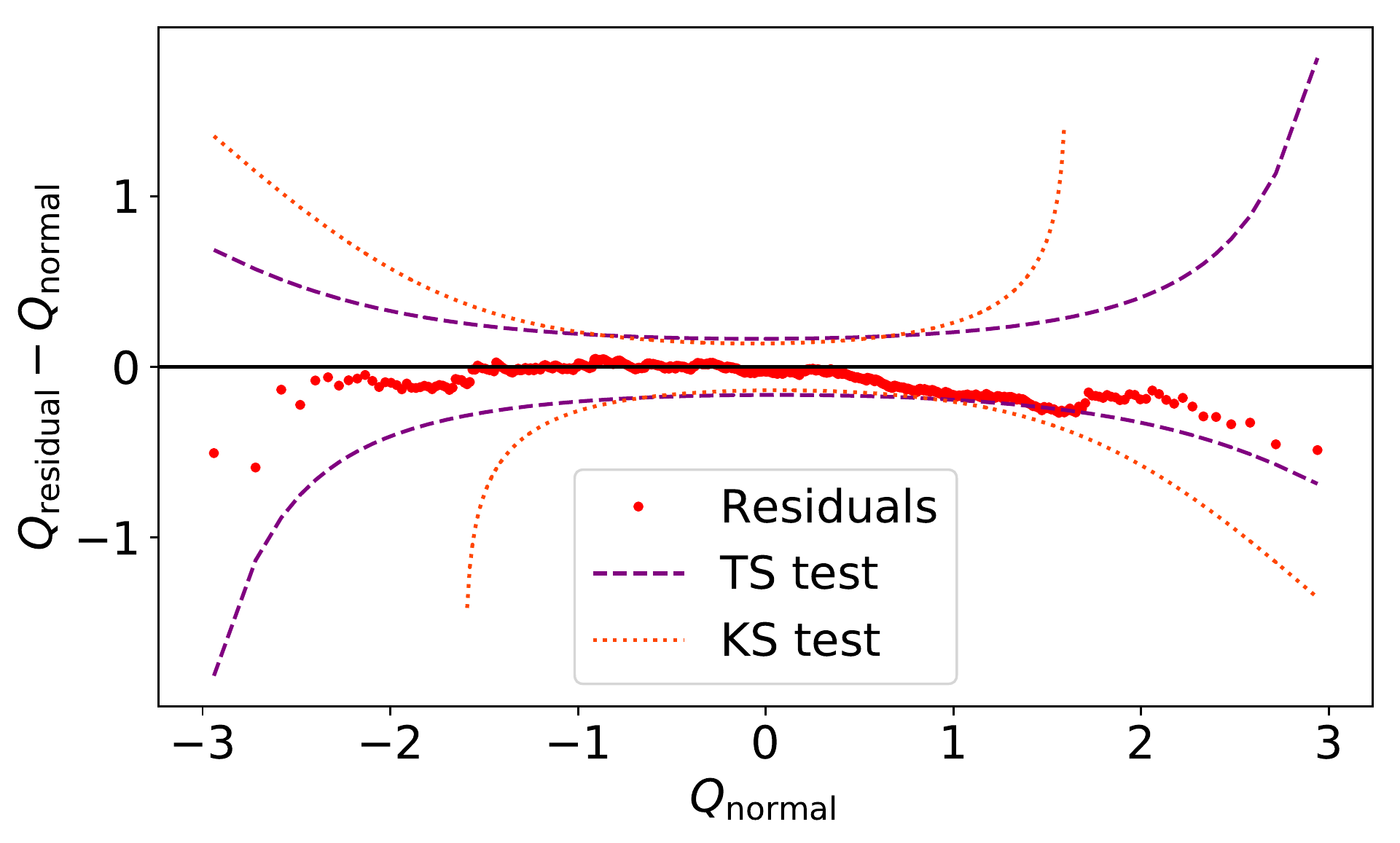}
    \caption{Rotated quantile-quantile (Q-Q) plot of energy residuals $(E_\alpha^{SM}(\bm{\lambda_\text{USDB}}) - E_\alpha^{exp})/\Delta E_\alpha$ with respect to standard normal distribution. The dashed and dotted lines in the Q-Q plot show the boundaries of TS and KS-tests respectively. Deviation from the horizontal axis indicates non-normal deviations in the data. The residual points crossing the dashed purple line around $Q_\text{normal}\approx1.5$ corresponds to the low $p$-value returned by the TS-test. (A brief explanation of Q-Q plots can be found in the Appendix.)}
    \label{fig:qqplot}
\end{figure}

Before proceeding with the sensitivity analysis, it is important to test the distribution of residuals $R_\alpha = (E_\alpha^{SM}(\bm{\lambda_\text{USDB}}) - E_\alpha^{exp})/\Delta E_\alpha$, 
shown in Fig.~\ref{fig:residuals},
since we will approximate it to be normally distributed (equivalent to Laplace's approximation discussed in Appendix A). We employ two statistical tests of normality: Kolmogorov-Smirnov \cite{kstest} (KS-test) and tail-sensitive \cite{navarro_perez_error_analysis_2015, aldor-noiman_Power_2013} (TS-test); the former is a typical test of overall normality, while the latter is more sensitive to features in the tails of the  distribution. Each test returns a $p$-value: we adopt the traditional significance threshold of $p>0.05$ as no significant evidence for deviations from the standard normal distribution. This is sometimes colloquially referred as agreement between the empirical and theoretical distributions. To visualize these tests of normality, we show a rotated quantile-quantile (Q-Q) plot of the residuals $R_\alpha$  in Fig. \ref{fig:qqplot}. The residuals appear to have a nearly normal distribution, and indeed the KS-test returns a $p$-value of 0.15. This validates our implementation of Laplace's approximation. However, the more sensitive TS-test returns a $p$-value of 0.02, indicating that the tails of the residual distribution contain sufficient non-normal features as to warrant a more detailed study in future work.

Under the assumption  the errors have a  normal distribution ,  $\chi^2$ is well-approximated by quadratic function in $\bm{\lambda}$, and we can compute the Hessian $H$, or the second derivative of $\chi^2$, that is,
\begin{equation}
H_{ij} = \frac{1}{2} \frac{\partial^2}{ \partial \lambda_i \partial \lambda_j}    \chi^2.
\end{equation}
Note that we write the Hessian matrix as $H$, and the Hamiltonian operator as $\hat{\mathcal{H}}$. 
We can simplify this expression to put it in terms of eigenenergies:
\begin{equation}
	\frac{\partial^2 \chi^2 
	}{\partial \lambda_i \partial \lambda_j} = \\
	\sum_{\alpha=1}^{N} \frac{2}{(\Delta E_\alpha)^2} \left[ \frac{\partial E^{SM}_\alpha}{\partial \lambda_i}  \frac{\partial E^{SM}_\alpha}{\partial \lambda_j} + (E^{SM}_\alpha - E_\alpha^{expt}) \frac{\partial^2 E^{SM}_\alpha}{\partial \lambda_i \partial \lambda_j} \right]
\end{equation}
so that
\begin{equation}
H_{ij} =  \sum_{\alpha=1}^{N} \frac{1}{(\Delta E_\alpha)^2} \frac{\partial E^{SM}_\alpha}{\partial \lambda_i} \frac{\partial E^{SM}_\alpha}{\partial \lambda_j} + \sum_{\alpha=1}^{N} \frac{(E^{SM}_\alpha - E_\alpha^{expt})}{(\Delta E_\alpha)^2} \frac{\partial^2 E^{SM}_\alpha}{\partial \lambda_i \partial \lambda_j}
\label{hessian}
\end{equation}
The first term in this expression dominates, so we define the \textit{approximate Hessian} $A$ as follows:
	\begin{equation}
H_{ij} \approx \sum_{\alpha=1}^{N} \frac{1}{(\Delta E_\alpha)^2} \frac{\partial E^{SM}_\alpha }{\partial \lambda_i} \frac{\partial E^{SM}_\alpha}{\partial \lambda_j} \equiv A_{ij}
\label{hess_approx}
\end{equation}
This approximation is good if the cross-derivative is small, for example if the energies were 
exactly linear in the parameters, or alternatively if the residual is small (meaning the model is good). Furthermore, the calculation of $E^{SM}_\alpha$ is made with the optimized USDB parameters, therefore the term multiplying the cross-derivative should on average be close to zero. The second term contains the cross-derivative, and this is more challenging to calculate, especially considering the size of the parameter-space. 

Note that the energy matrix element is nonlinear in $\bm{\lambda}$ due to dependence in the wavefunction. If one were to ignore this dependence, we call this the \textit{linear model approximation}
\begin{equation}
    E  = \langle \psi (\bm{\lambda}) | \hat{\cal H} (\bm{\lambda}) | \psi (\bm{\lambda}) \rangle \approx \langle \psi | \hat{\cal H} (\bm{\lambda}) | \psi \rangle
     = \sum_{i=1} \lambda_i \langle \psi | \hat{\cal O}_i | \psi \rangle
    \label{eqn:lma}
\end{equation}
Under the linear model approximation, any cross-derivative term is zero and thus the `approximate' Hessian above would be  equal to the full Hessian: $A = H$.


To compute the derivatives of the  energies, in Eq.\ref{hess_approx}, 
we use the Feynman-Hellmann theorem,
\begin{equation}
\frac{\partial E^{SM}_\alpha(\bm{\lambda})}{\partial \lambda_i} = \left \langle \psi_\alpha \left | \frac{d\hat{\mathcal{H}}}{d\lambda_i} \right | \psi_\alpha \right \rangle
=  \langle \psi_\alpha | \hat{\mathcal{O}}_i | \psi_\alpha \rangle,
\end{equation}
where the Hamiltonian (\ref{Hamiltonian_as_sum}) is linear in $ \lambda_i $.
(These first derivatives are Jacobians \cite{dobaczewski2014error}.)
Thus, for the first derivatives in (\ref{hess_approx}), we can simply evaluate expectation values of the individual 1- and 2-body operators.

While the full numerical calculation of the Hessian is quite costly, we can numerically compute the cross-derivative term in Eq.~\ref{hessian} with a simple finite difference approximation of the second derivative, so as to achieve a better approximation to the exact Hessian.

\begin{equation}
 \frac{\partial^2 E^{SM}_\alpha }{\partial \lambda_i \partial \lambda_j} = \frac{1}{2 \epsilon} \left[  \frac{\partial E^{SM}_\alpha(\bm{\lambda}_j^+)}{\partial \lambda_i} - \frac{\partial E^{SM}_\alpha(\bm{\lambda}_j^-)}{\partial \lambda_i}  \right] + \mathcal{O}(\epsilon^2)
\end{equation}
 
 Here, $E_\alpha^{SM}(\bm{\lambda}_j^\pm)$ is the $\alpha$-th energy evaluated using USDB parameters with the $j$-th value perturbed by $\pm \epsilon$. Inserting into Eq.~\ref{hessian}, we denote the resulting numerically corrected approximate Hessian matrix as $A_\text{num}$.
 \begin{equation}
     \left[A_\text{num}\right]_{ij} \equiv [A]_{ij} + \sum_{\alpha=1}^{N} \frac{(E^{SM}_\alpha(\bm{\lambda}) - E_\alpha^{expt})}{(\Delta E_\alpha)^2} \frac{1}{2 \epsilon} \left[  \frac{\partial E^{SM}_\alpha(\bm{\lambda_j^+})}{\partial \lambda_i} - \frac{\partial E^{SM}_\alpha(\bm{\lambda_j^-})}{\partial \lambda_i}  \right]
 \end{equation}{}

 \begin{table}[]
    \centering
    \begin{tabular}{|c|c|c|c|c|}
    \hline
         $i$ & $[\Delta]_{ii}$ & $\sigma_i$  & $[\Delta_\mathrm{num}]_{ii}$ & $[\sigma_\mathrm{num} ]_i$  \\
           &  (MeV$^{-2}$)  & (keV)    &  (MeV$^{-2}$) & (keV) \\
         \hline
         1  &   11785000     & $0.29$ &  11785500 &   $0.29$               \\
         2  &  393000        & 1.6  &   393600    &   1.6            \\ 
         3  &   79100        & 3.5  &   78810     &   3.5            \\ 
         4  &   71200        & 3.7  &   70800     &   3.7           \\ 
         5  &  22200         & 6.7  &   22220     &   6.7           \\ 
         6  &   6357         & 13 &   6357      &   13          \\ 
         7  &   5200         & 14 &   5175      &   14          \\ 
         8  &   3600         & 17 &   3590      &   17          \\ 
         9  &   3270         & 17 &   3261      &   17            \\ 
         10 &   3050         & 18 &   3035      &   18           \\ 
         ... &  ...          & ... & ...        &   ...         \\ 
         64  &  10.6          &  307 &   $<1$     &    $>1000$           \\
         65  &  7.71          &  360 &   $<0.1$   &   $>3000$             \\
         66  &  3.16          &  562 &    $<0.1$  &   $>3000$             \\
         
         \hline
         
    \end{tabular}
    \caption{Statistics of linear-combinations of USDB matrix elements, or \textit{principle component analysis} (PCA) parameters. The eigenvalues of the approximate Hessian matrix $A$ we denote as $[\Delta]_{ii}$, 
    which is the  sensitivity of the $i$th PCA parameter, and $\sigma_i$ is the corresponding uncertainty. Thus the most sensitive 
    PCA parameter is constrained to within 290 eV. Likewise, the eigenvalues of the numerically corrected approximate Hessian matrix $A_\text{num}$ we denote as $[\Delta_\text{num}]_{ii}$, and $[\sigma_\text{num}]_i$ is the corresponding uncertainty. Note that for the most sensitive PCA parameters, the numerical correction effectively leaves the standard deviations unchanged.}
    \label{tab:milcom_table}
\end{table}

We tested their importance by evaluating with $\epsilon \approx 0.1$. The  
the resulting eigenvalues of $A$ and $A_\text{num}$, shown in Table \ref{tab:milcom_table}, 
are very similar, indicating that while
 the numerical corrections terms are individually nonzero, the total contributions  average to very small contributions.
Thus $A$ is in fact a very good approximation to the full Hessian matrix and, in what follows, we find that propagation of 
uncertainties onto observables are almost independent of the numerical correction. This also implies that the linear model approximation (Eq. \ref{eqn:lma}) is a good approximation.

\subsection{PCA Transformation}

The Hessian, whether exact ($H$) or approximate ($A$), allows us to determine the 
uncertainty in parameters, discussed in more detail in section \ref{uncertain}, and in particular the uncorrelated uncertainties. 
Transforming the Hessian $ U H U^T = D$, where $D$ is diagonal, or its approximation 
\begin{equation}
 W A W^T   = \Delta
\end{equation}
where $\Delta \approx D$ is also diagonal,
provides a transformation from the original parameters $ \bm{\lambda}$ to 
new linear combinations of parameters,
\begin{equation}
\bm{\mu}= W \bm{\lambda}.    \label{PCAparam}
\end{equation}
This is simply \textit{principal component analysis} (PCA) of the  Hessian, and so we call  $\bm{\mu}$ the 
PCA parameters.  In terms of our approximate Hessian, we can also understand this as 
a singular value decomposition (SVD) of the linear response $J_{\alpha i } =\partial E_\alpha / \partial \lambda_i$. More formally, we approximate $H \approx A = J^T \Sigma^{-2} J$,
where $\Sigma$ is the diagonal matrix of 
uncertainties on energies, $\Sigma_{\alpha \beta} = \delta_{\alpha \beta} \Delta E_\alpha$; but, as is nearly true, $\Delta E_\alpha 
\approx \Delta E^{th}$ and hence $A \approx (\Delta E^{th})^{-2} J^T J$; then it should 
be clear that the eigenvalues of $A$ are proportional to the SVD eigenvalues of $J$. 
Thus the eigenvalues found in $\Delta$, presented in Table \ref{tab:milcom_table} and plotted in Fig.~\ref{milcom_plot}, allow us to determine the 
most important linear combinations of parameters to the fit.

 \section{Evaluating uncertainties}
 \label{uncertain}
 \begin{figure}
    \centering
    \includegraphics[scale=1,clip]{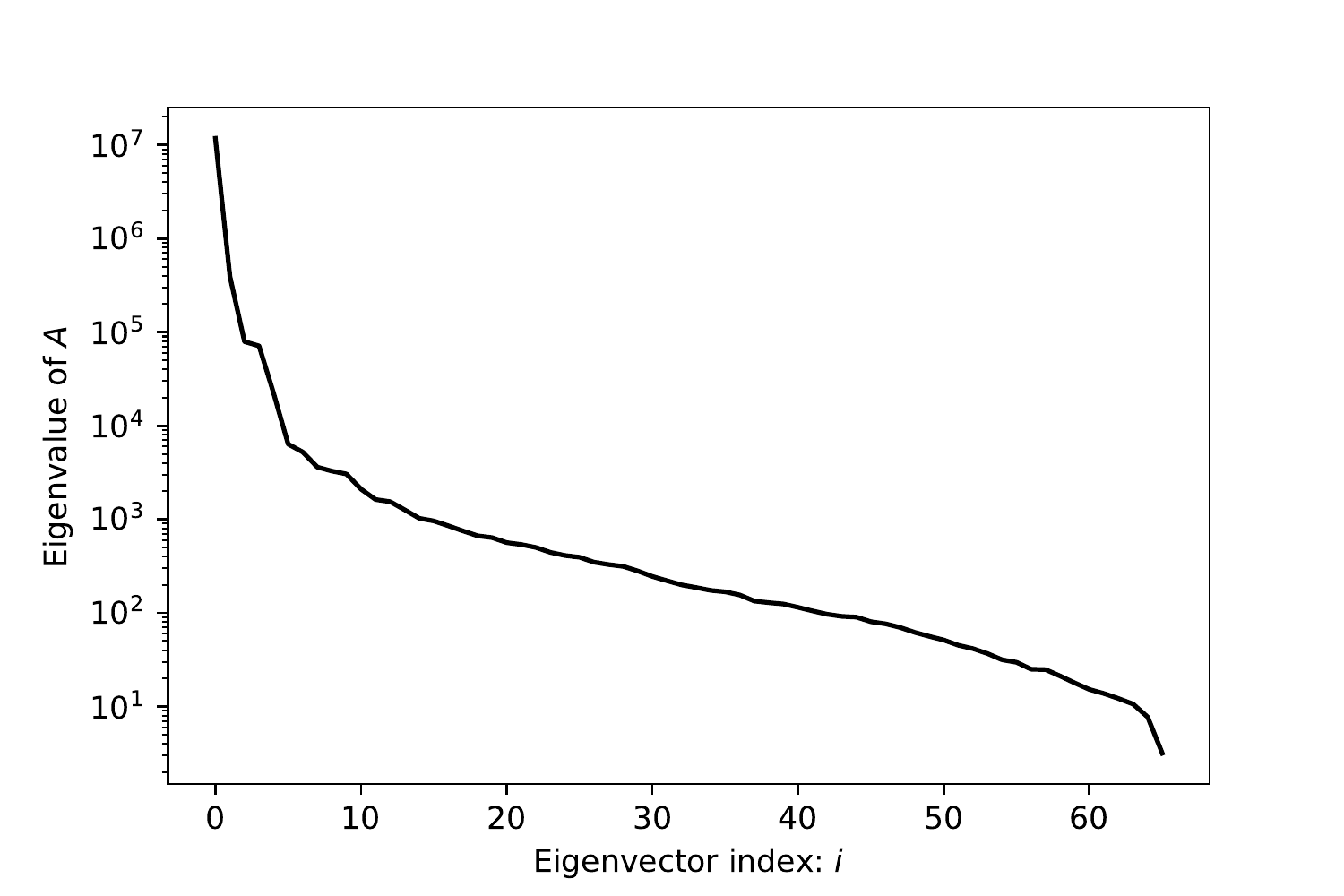}
    \caption{Ordered eigenvalues of the approximate Hessian $A$, which equal the diagonal elements of $\Delta$. The eigenvalues are interpreted as the sensitivity of the corresponding linear combination or 
    principal components of matrix elements (PCA-parameter). The first PCA-parameter carries 95\% of the total sensitivity, and the first 5 PCA-parameters carry 99.6\% of the sensitivity. }
    \label{milcom_plot}
\end{figure}

 The parameter covariance matrix is simply the inverse of the Hessian matrix, which we 
 have approximated as
 \begin{equation}
    	C(\bm{\lambda}) = H^{-1}\approx A^{-1} = W^{T}\Delta^{-1} W
 \label{covariance}
 \end{equation}
 The naive variance of  the original parameters $\mathbf{\lambda}$ is given by the diagonals of the covariance matrix, so that  $\sigma(\lambda_i) = \sqrt{C_{ii}}$. 
 This, however,  ignores correlations between parameters and thus is an  incomplete description of parameter 
 uncertainties. A better approach is to compute variances from the diagonalized Hessian matrix, and thus obtaining uncorrelated 
 uncertainties on the PCA parameters, 
  $\sigma({\bm \mu}) = 1/\sqrt{\Delta_{ii}}$.
  These we give in Table \ref{tab:milcom_table}, and plot
  in Fig. \ref{milcom_plot}. Here one sees the first few PCA parameters have very large sensitivity, and indeed the first 10 carry over 99.8\% of the total; it is well-known lore in the nuclear shell-model community the fit of USDB and similar empirical interactions are dominated by only
  a few linear combinations, which here define the PCA parameters. Table~\ref{tab:milcom_table} in fact demonstrates 
  these parameters must be known to within a few keV or better; on the other hand 23 PCA parameters have uncertainties of 100-500 keV. At this point, it is important to remember that these variablities are with respect to experimental data that only includes energies, so these low-variability PCA parameters could in principle be tuned to fit the interaction to various other observables without disrupting the fit to energies.

If the 
uncertainties in the principal components ${\bm \mu}$ are independent, then the propagation of 
uncertainties is straightforward. For 
any observable $X$,
\begin{equation}
    \sigma^2(X) = \sum_i \left (\frac{\partial X}{\partial \mu_i} \right )^2\sigma^2(\mu_i)
\end{equation}
Using (\ref{PCAparam}), 
\begin{equation}
\frac{\partial X}{\partial \mu_i}= \sum_j W_{ij} \frac{\partial X}{\partial \lambda_j}
\end{equation}
and so
\begin{equation}
    \sigma^2(X) = \sum_i\sigma^2(\mu_i) \sum_{jk} \frac{\partial X}{\partial \lambda_j} W_{ij}
    \frac{\partial X}{\partial \lambda_k}W_{ik} 
    = \bm{g}^T C \bm{g}
    \label{sigma_obs}
\end{equation}
where $g_i = \partial X /\partial \lambda_i$ is the linear response of any observable to 
perturbations in the original parameters. This is particularly useful in the case of 
energies, where we already have the linear response, thanks to the Feynman-Hellmann theorem.
For a discussion of some of the subtleties, see Appendix B.


For other observables, we do not use (\ref{sigma_obs}) directly. 
Instead, we generate perturbations in USDB by generating perturbations in the PCA parameters 
$\delta\bm{\mu}$ with a Gaussian distribution with width $\sigma(\mu_i)$ given by Table I.
Because the 
uncertainties are independent, or nearly so, in the PCA parameter representation, it is 
safe to generate the perturbations independently. We then transform back to the original 
representation of the matrix elements and read into a shell-model code \cite{BIGSTICK,johnson2018bigstick}, find eigenpairs,  and evaluate the reduced transition matrix element for one-body transition operators. We sampled at least $1,000$ sets of parameters, which gives sufficient convergence of the resulting set of matrix elements: assuming the transition strengths $B_i$ are normally distributed with respect to small perturbations in the Hamiltonian, we take the theoretical uncertainty $\sigma(B_i) $ as equal to the standard deviation of the set of samples.
Previous works have demonstrated convergence with similar approaches and an even smaller number of samples. In \cite{Perez:2014laa} the statistical uncertainty in the binding energy of $^3$H was quantified using $250$ samples of an interaction with about $40$ parameters, resulting in $\sigma(B) = 15$ keV. The same result was later reproduced in \cite{Perez:2015bqa} using only $33$ samples.

\begin{figure}
    \includegraphics[scale=0.9,clip]{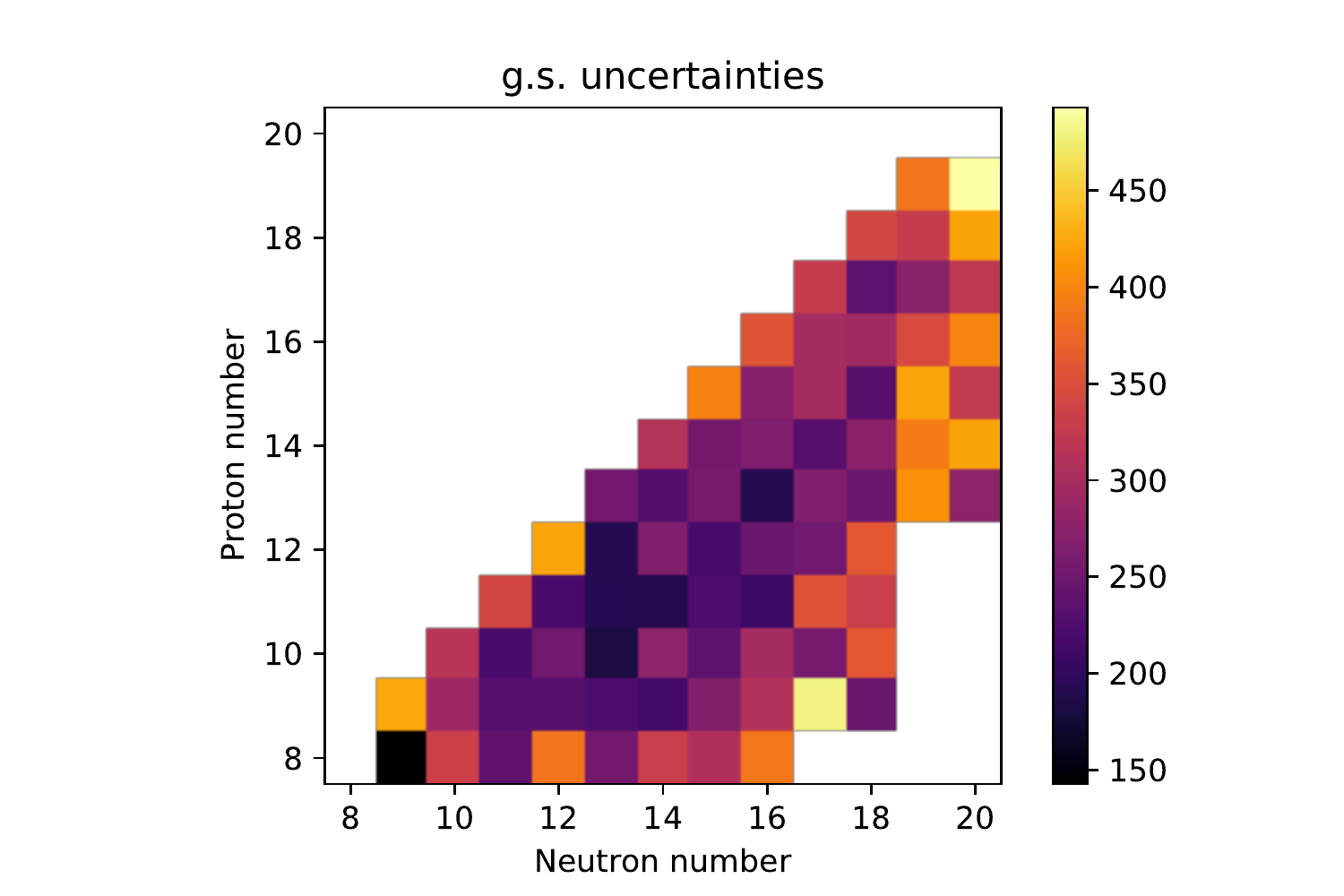} 
\caption{Estimated $1\sigma$ uncertainties of ground-state fit energies in units of keV.} 
\label{fig:gs_uncertainty}
\end{figure}

 \subsection{Results}
 
 
 For the energies used in the fit, we already have the elements of $ \vec{g} $ saved from computing the approximate Hessian, so this calculation is cheap. We can thus estimate covariance in the computed energies $C_E$ by expanding this expression to a matrix equation.
 \begin{equation}
 	C_E = J C_\lambda J^T
 	\label{energy_cov}
 \end{equation}
 
 Results for some of these estimated uncertainties are given in Table \ref{tab:fit_errors}. Using these estimates, 75\% of shell-model energies are within $1\sigma$ of experiment, and 96\% are within $3\sigma$; these are close to the standard normal quantiles of 68\% and 99\% respectively, so we conclude that these theoretical uncertainties are sensible. Akin to the original 
 sensitivity analysis of fit energies \cite{PhysRevC.74.034315}, Fig. \ref{fig:gs_uncertainty} shows theoretical $1\sigma$ uncertainties on ground-state binding energies. We refer the reader to \cite{PhysRevC.74.034315} for comparison to 
 uncertainty plots, in particular Fig. 10 of that paper. While this description of 
 uncertainties on the fitted energies may be useful, we also note that they are in a  sense tautological: the energy covariance $C_E$ is related to the energy 
 uncertainties in Eq. \ref{add_errors} by a coordinate transformation. An algebraic explanation  is given in Appendix B.

 \begin{table}[]
    \centering
    \begin{tabular}{|c|c|c|c|c|c|}
    \hline
         Nucleus & $J^\pi_n$ & $T$ & $E^{exp}-E^{SM}$  & $\sigma$  \\
                 &           &     &     (keV)       &   (keV) \\
         \hline
        $^{30}$Si & $1^+_1$ & 1 &           -114 &      851\\
        $^{39}$K & $1/2^+_1$ & 1/2 &     -189 &      785\\
        $^{25}$F & $5/2^+_1$ & 7/2 &     -312.1 &    743\\
        $^{38}$K & $1^+_1$ & 0 &         -355.9 &    686\\
        $^{27}$Al & $11/2^+_1$ & 1/2 &  -52.9 &     615\\
        ...&...&...&...&...\\
        $^{24}$Mg & $6^+_1$ & 0 &       156.1 &     156\\
        $^{20}$Ne & $6^+_1$ & 0 &       -223.2 &    154\\
        $^{23}$Na & $11/2^+_1$ & 1/2 &  -15.3 &     153\\
        $^{28}$Mg & $2^+_1$ & 2 &       19.3 &      153\\
        $^{17}$O & $5/2^+_1$ & 1/2 &    218.3 &     142\\
        \hline
         
    \end{tabular}
    \caption{States in experimental energy data, shown in order of descending uncertainty $\sigma$ (high-variability on top, low-variability on bottom). }
    \label{tab:fit_errors}
\end{table}

\begin{figure}
\begin{tabular}{cc}
  \includegraphics[scale=0.5,clip]{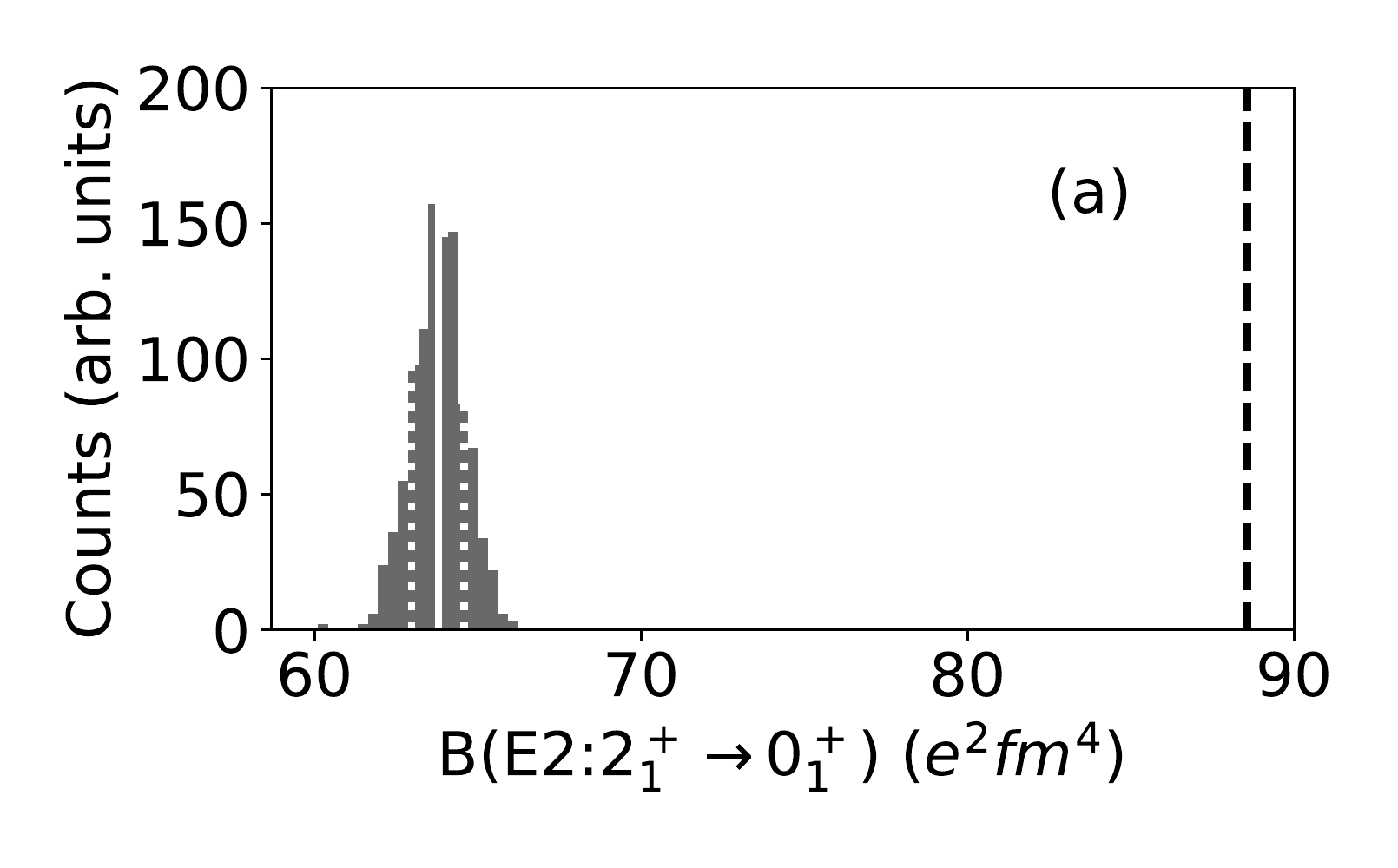} &   \includegraphics[scale=0.5,clip]{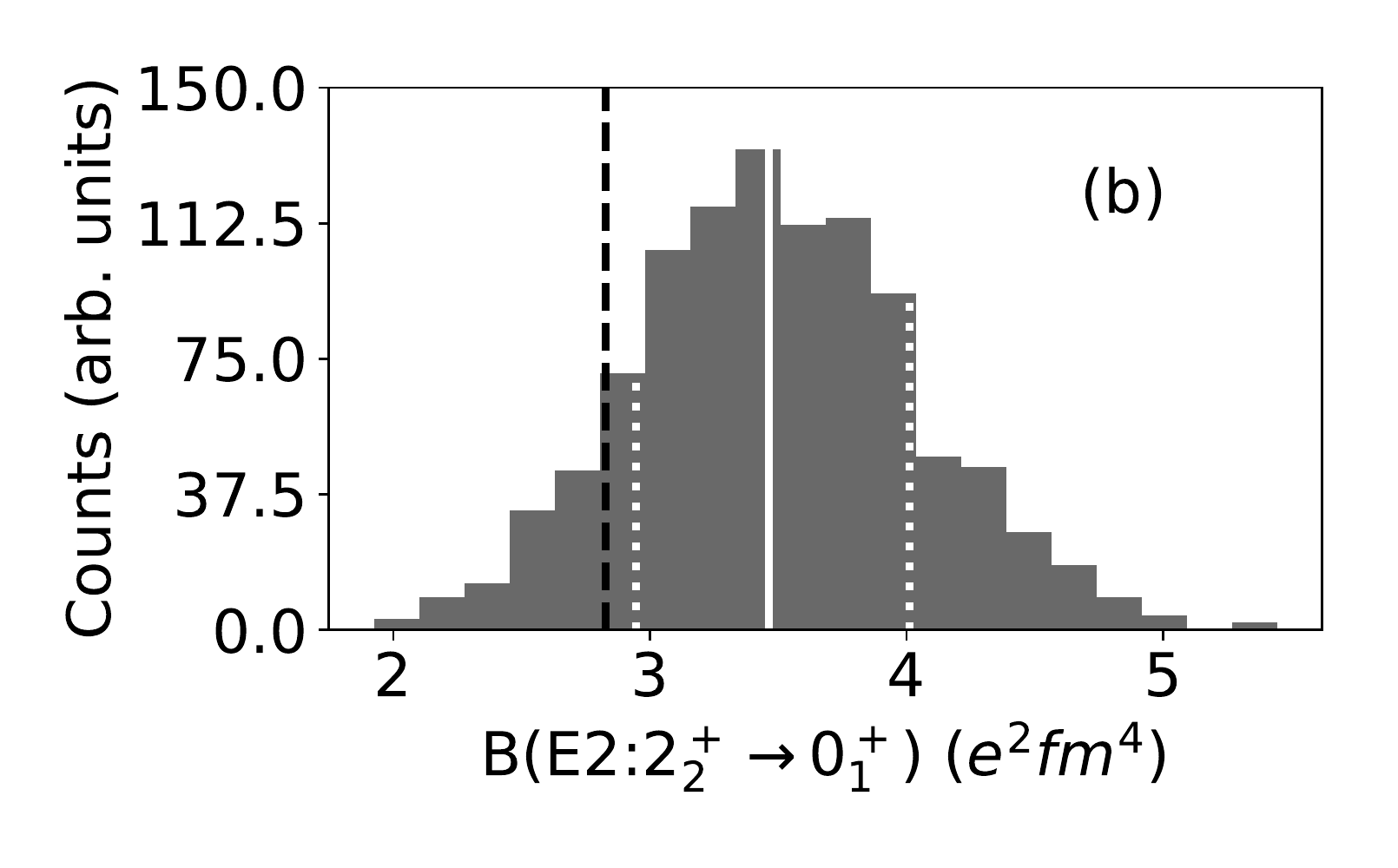} \\
 \includegraphics[scale=0.5,clip]{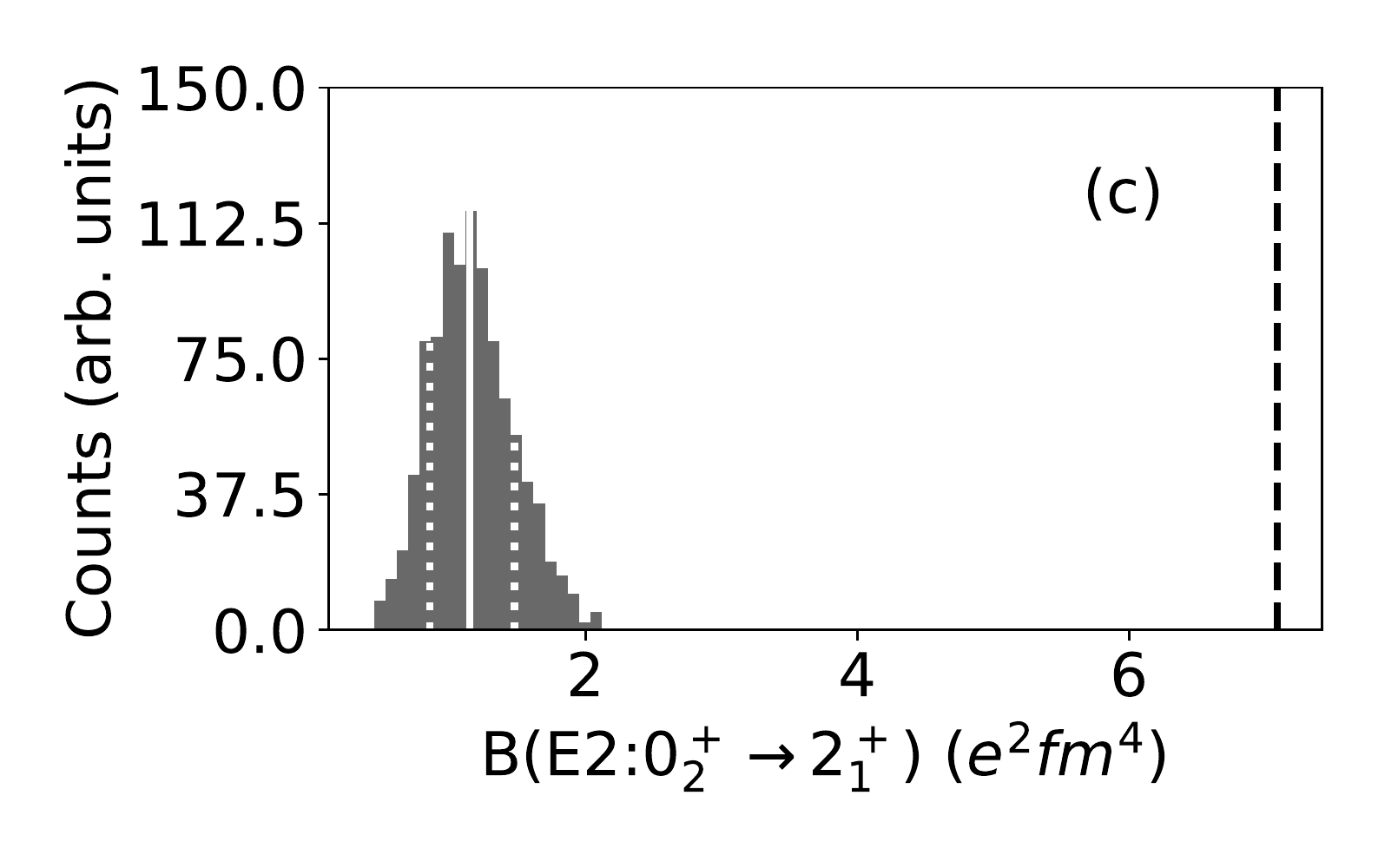} &   \includegraphics[scale=0.5,clip]{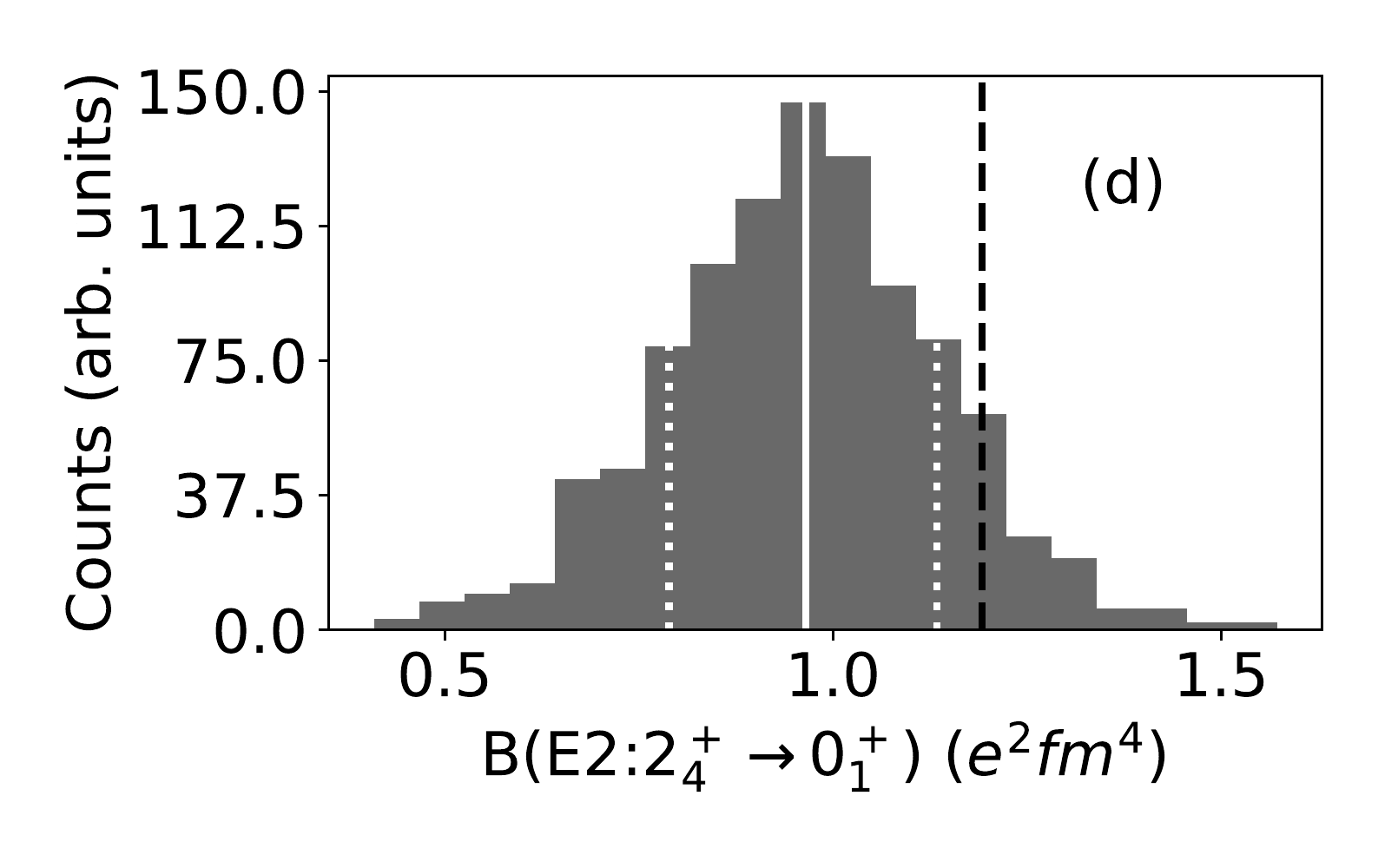} \\
\end{tabular}
\caption{Distributions of the electric quadrupole (E2) transition strengths for $^{26}$Mg.   Black dashed line shows experimental value \cite{BASUNIA20161}. 
The the median values and uncertainty interval are highlighted in white:
 (a) $2^+_1 \rightarrow 0^+_1:63.7^{+0.78}_ {-0.83}$, 
 (b) $2^+_2 \rightarrow 0^+_1:3.46^{ +0.55}_{{-0.52 }}$ , 
 (c) $0^+_2 \rightarrow 2^+_1: 1.15^{ +0.33}_{ -0.29 }$ , (d) $2^+_4 \rightarrow 0^+_1: 0.96^{ +0.18}_{ -0.18 }$, all in units of $\text{e}^2\text{fm}^4$.
   }



\label{fig:E2Mg26}
\end{figure}

 \begin{figure}
\begin{tabular}{cc}
  \includegraphics[scale=0.5,clip]{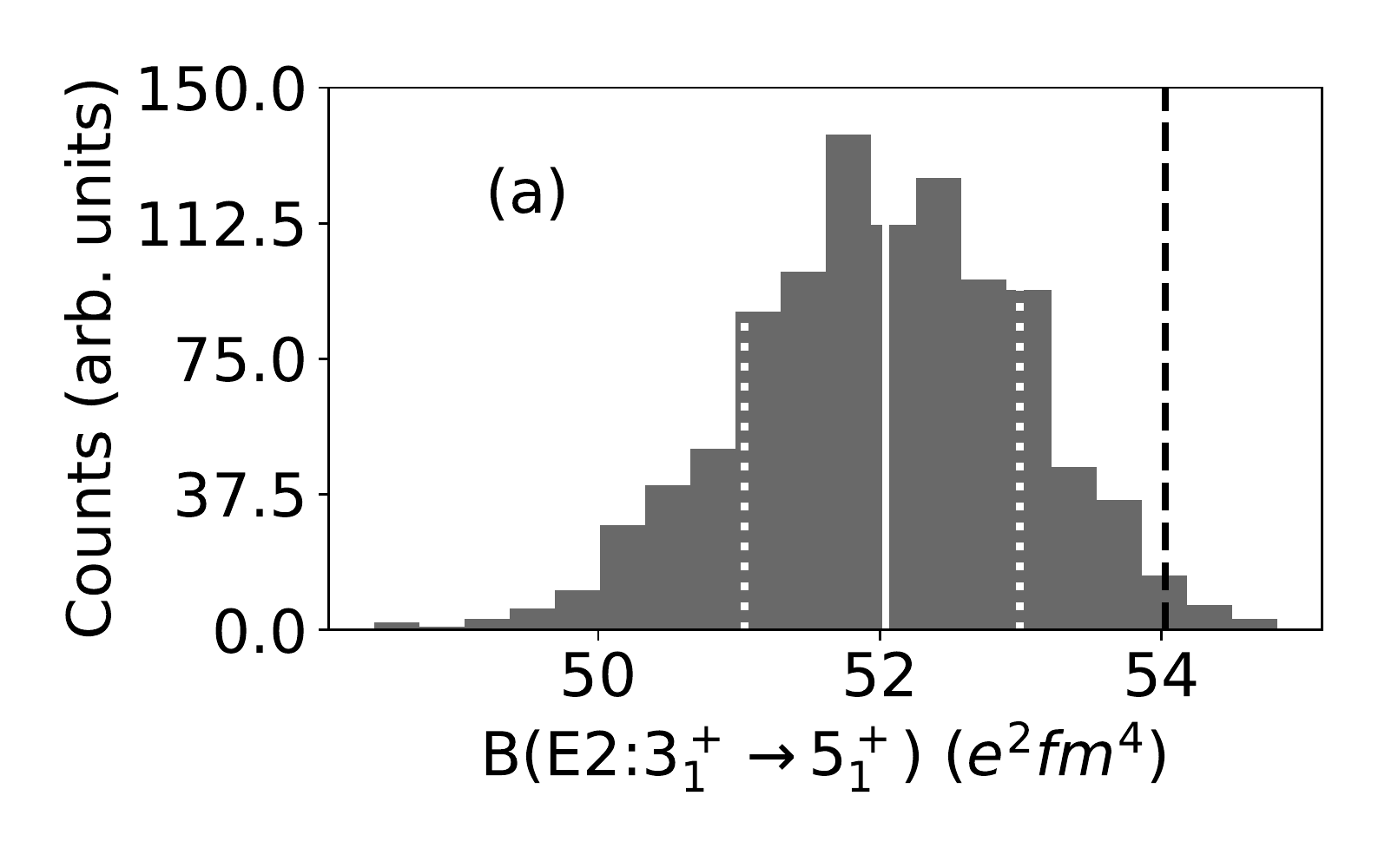} &   \includegraphics[scale=0.5,clip]{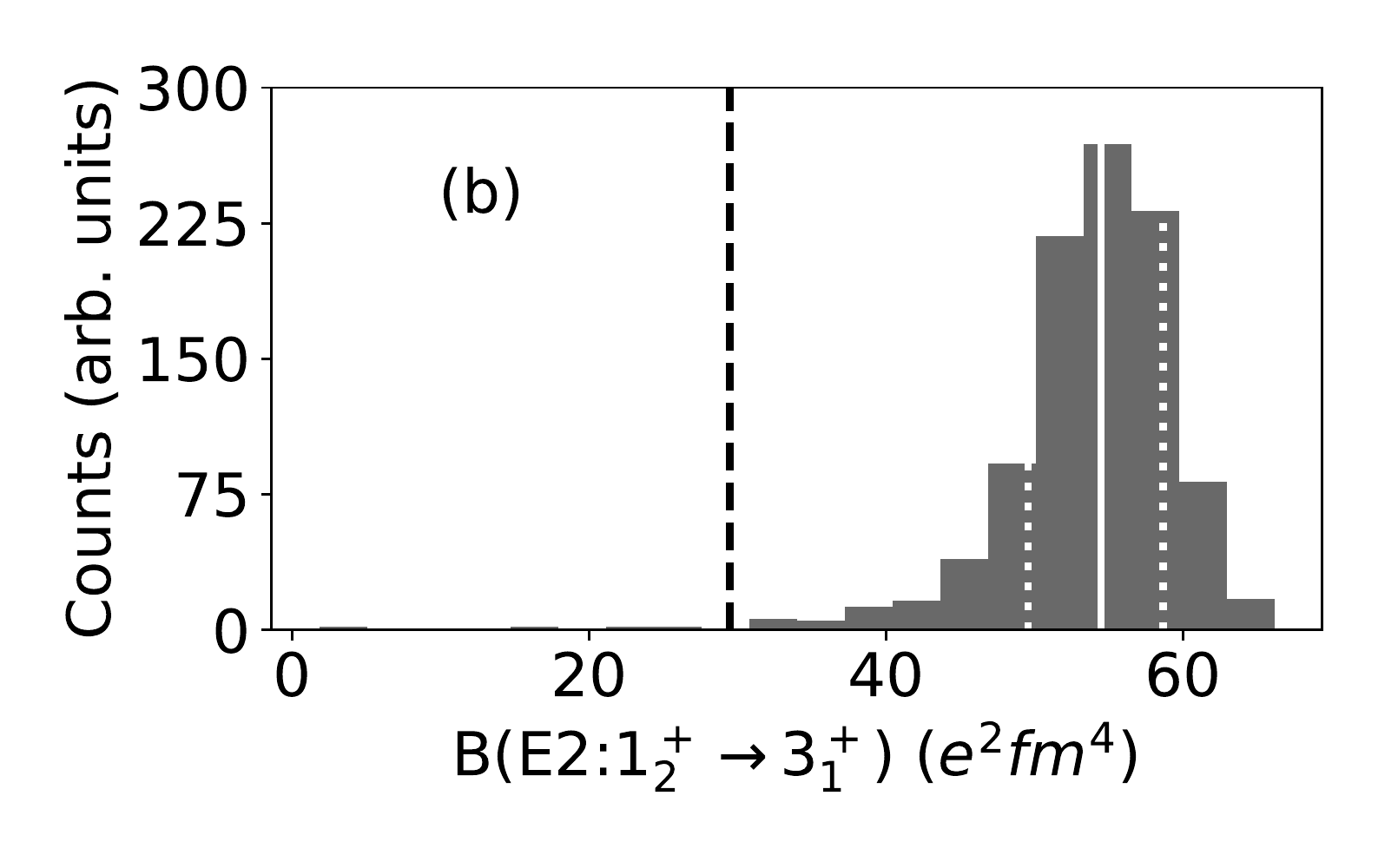} \\
 \includegraphics[scale=0.5,clip]{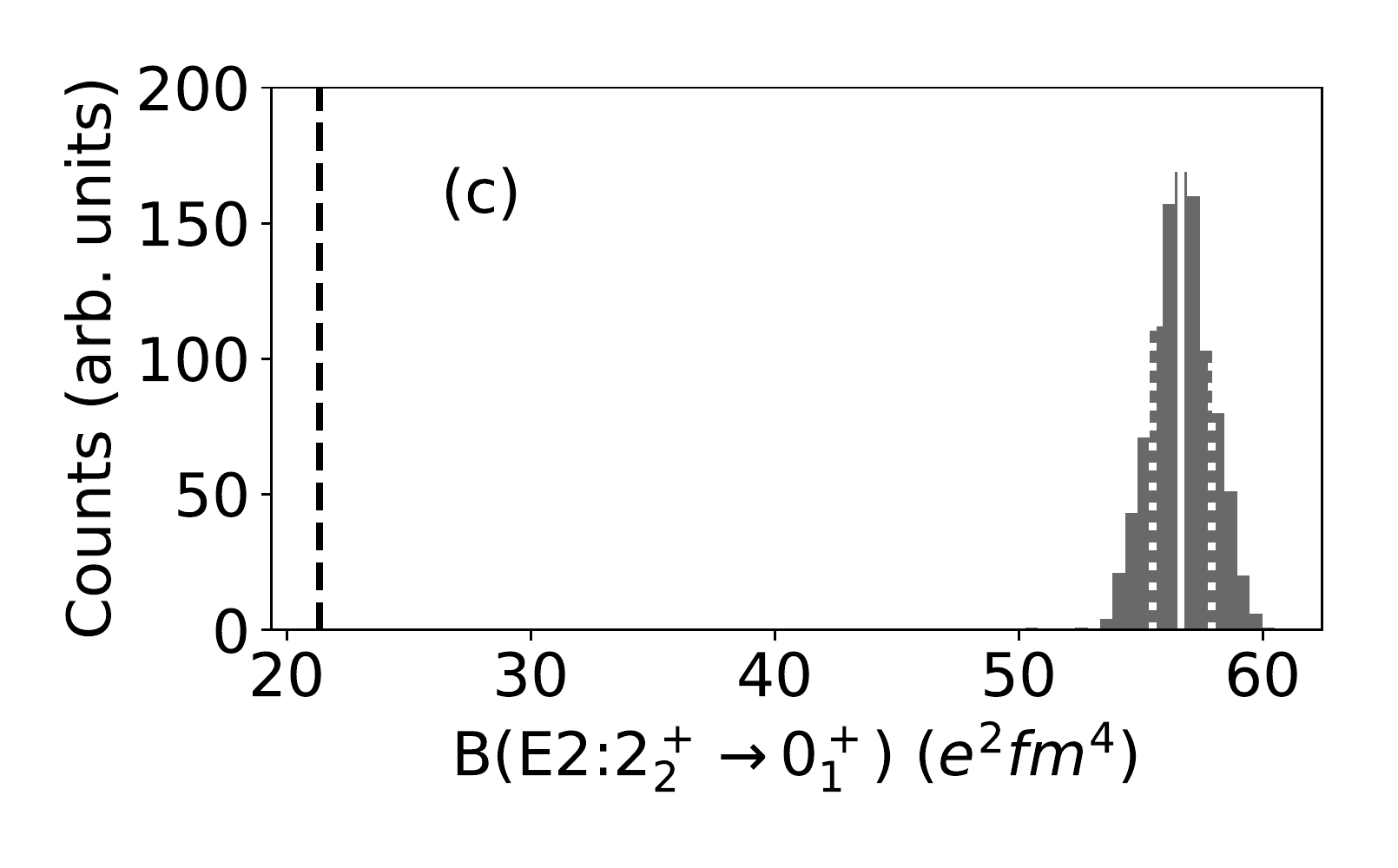} &   \includegraphics[scale=0.5,clip]{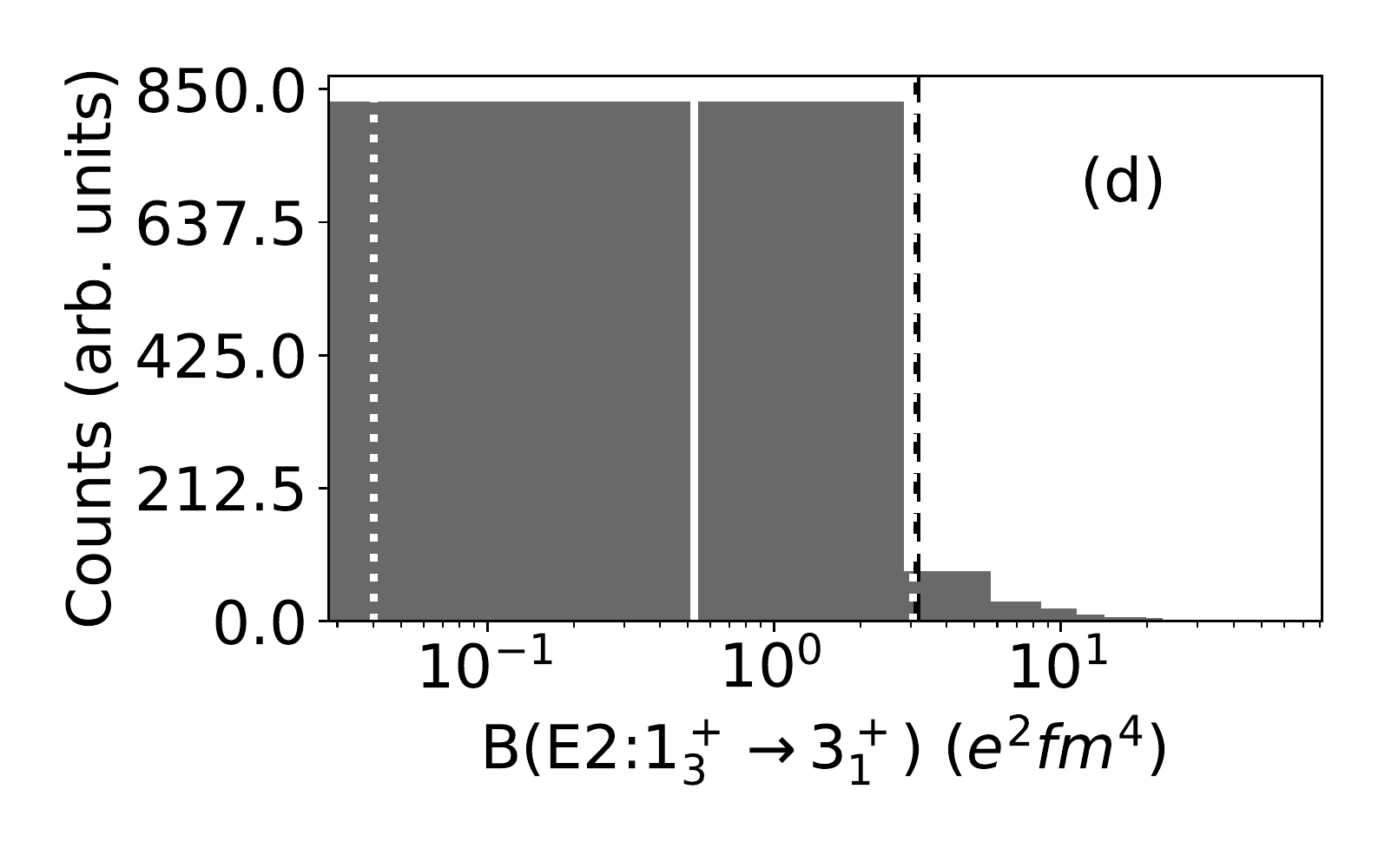} \\
 \includegraphics[scale=0.5,clip]{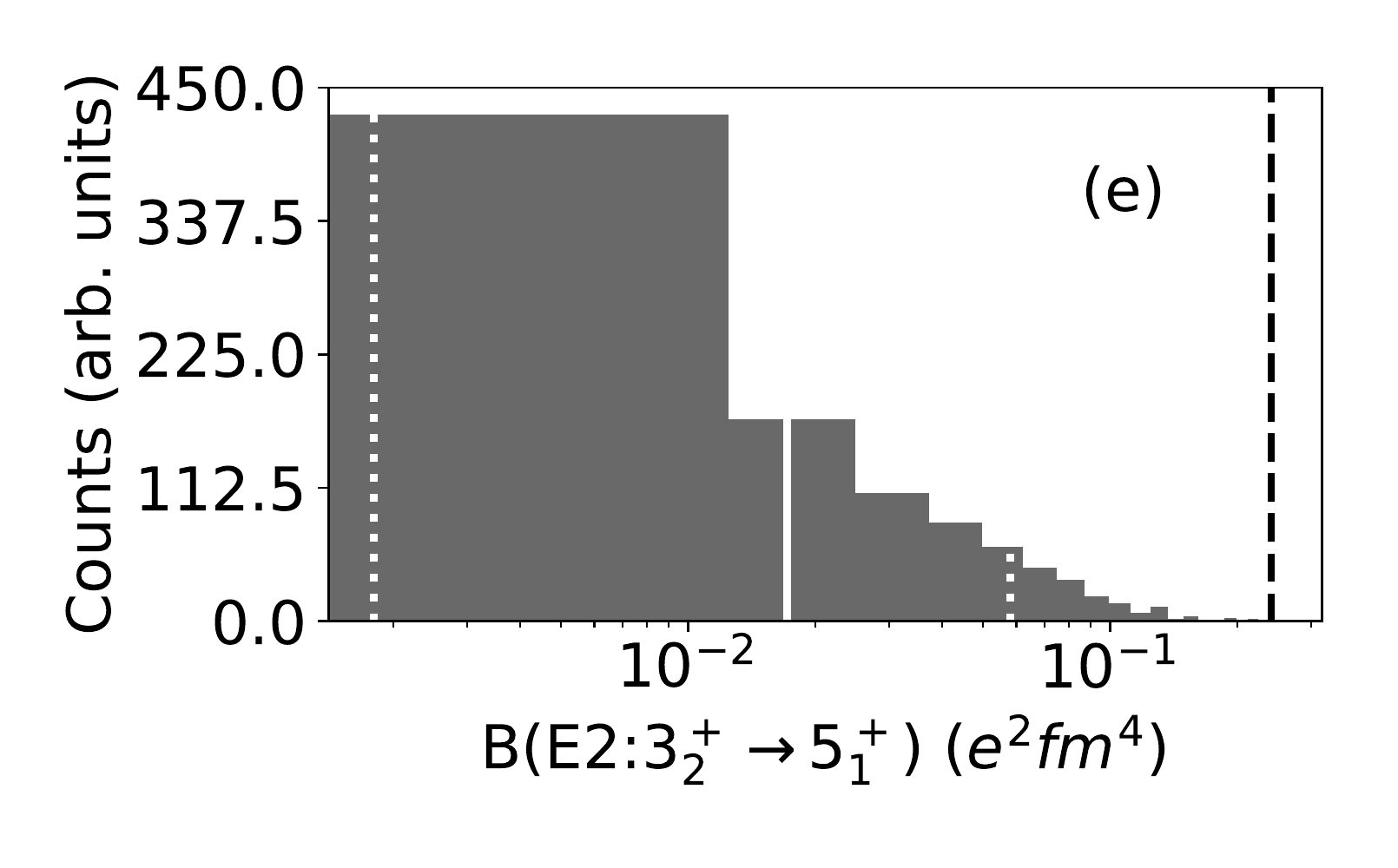} &   \includegraphics[scale=0.5,clip]{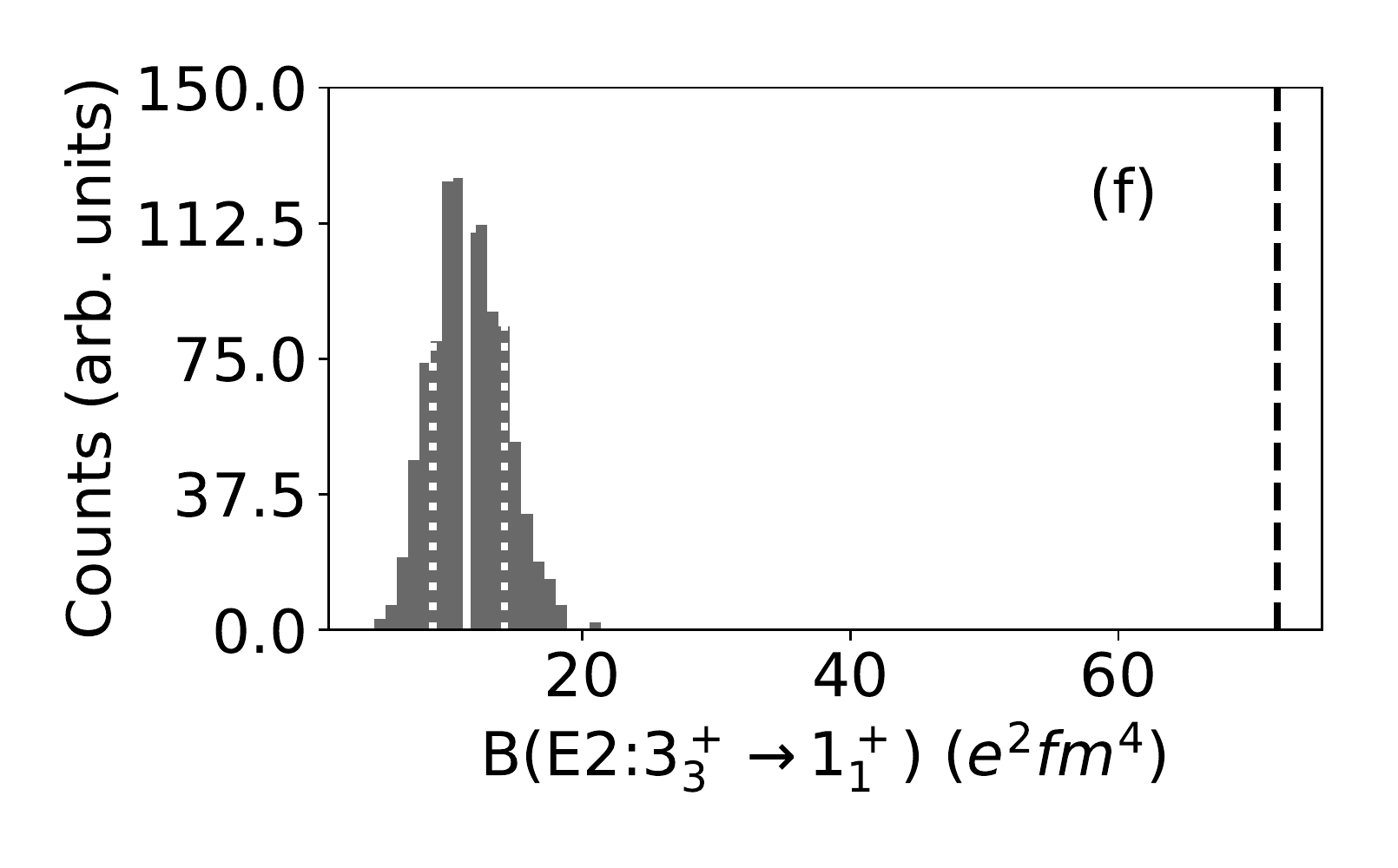} \\
\end{tabular}
\caption{Distributions of the electric quadrupole (E2) transition strengths for $^{26}$Al.   Black dashed line shows experimental value \cite{BASUNIA20161}. The median values and uncertainty intervals are highlighted in white : 
(a) $3^+_1 \rightarrow 5^+_1:$ 
$52.04^{ +0.99}_{{ -1.0 }}$,
(b) $1^+_2 \rightarrow 3^+_1:$
$54.47^{ +4.19}_{ -4.92 }$ , 
(c) $2^+_2 \rightarrow 0^+_1:$
$56.63^{ +1.26}_{ -1.16 }$ ,
(d) $1^+_3 \rightarrow 3^+_1:$
$0.53^{ +2.53}_{ -0.49 }$, 
(e) $3^+_2 \rightarrow 5^+_1:$
$0.017^{ +0.041}_{ -0.015 }$, and 
(f) $3^+_3 \rightarrow 1^+_1:$
$11.38^{ +2.82}_{ -2.53 }$, all in units of $\text{e}^2\text{fm}^4$ .
}





\label{fig:E2Al26}
\end{figure}

    We also computed the uncertainties in  selected transitions. The uncertainty bands presented in all transition strength calculations correspond to the 16th - 84th percentiles; for normal distributions this is precisely the $1\sigma$ uncertainty band, but we find many computed transition strengths have asymmetric distributions (especially those with small B-values). This, along with reporting median rather than mean, gives a more accurate description of uncertainty.
    
    Following \cite{PhysRevC.80.034301}, 
    we compute reduced transition strengths B(E2) for several low-lying transitions in $^{26}$Mg and $^{26}$Al, shown in Fig. \ref{fig:E2Mg26} and \ref{fig:E2Al26} respectively. The one-body electric quadrupole operator matrix elements were computed assuming
    harmonic oscillator radial wave functions  with oscillator length $b=1.802$
    \cite{br88} and effective charges  $e_p = 1.36$, $e_n = 0.45$, which were obtained by a least-squares fit \cite{PhysRevC.78.064302}.  While some values are close to experiment, others differ significantly. The B(E2) values are quadratically dependent upon both the oscillator length and the effective charges, and can 
    be quite sensitive to small changes in the interaction matrix 
    elements \cite{PhysRevC.80.034301}.

    For $^{26}$Mg, in Fig.~\ref{fig:E2Mg26}, the median values and 
uncertainty intervals for our selected transitions are 
$2^+_1 \rightarrow 0^+_1:$ $63.7 \substack{+0.78 \\ -0.83}$, $2^+_2 \rightarrow 0^+_1:$$3.46 \substack{ +0.55 \\ -0.52 }$ , $0^+_2 \rightarrow 2^+_1:$$1.15 \substack{ +0.33 \\ -0.29 }$ , and $2^+_4 \rightarrow 0^+_1:$ $0.96 \substack{ +0.18 \\ -0.18 }$, all in units of $\text{e}^2\text{fm}^4$,
while for $^{26}$Al, in Fig.~\ref{fig:E2Al26} the median values 
and uncertainty intervals for our selected transitions are
$3^+_1 \rightarrow 5^+_1:$ $52.04 \substack{ +0.99 \\ -1.0 }$,
$1^+_2 \rightarrow 3^+_1:$
$54.47^{ +4.19 }_{ -4.92 }$ , 
$2^+_2 \rightarrow 0^+_1:$
$56.63 ^{ +1.26 }_{ -1.16 }$ ,
$1^+_3 \rightarrow 3^+_1:$
$0.53 ^{ +2.53 }_{ -0.49 }$, 
$3^+_2 \rightarrow 5^+_1:$
$0.017 ^{ +0.041}_{ -0.015 }$, and 
$3^+_3 \rightarrow 1^+_1:$
$11.38 ^{ +2.82}_{ -2.53 }$.

\begin{figure}
    \begin{tabular}{cc}
      \includegraphics[scale=0.5,clip]{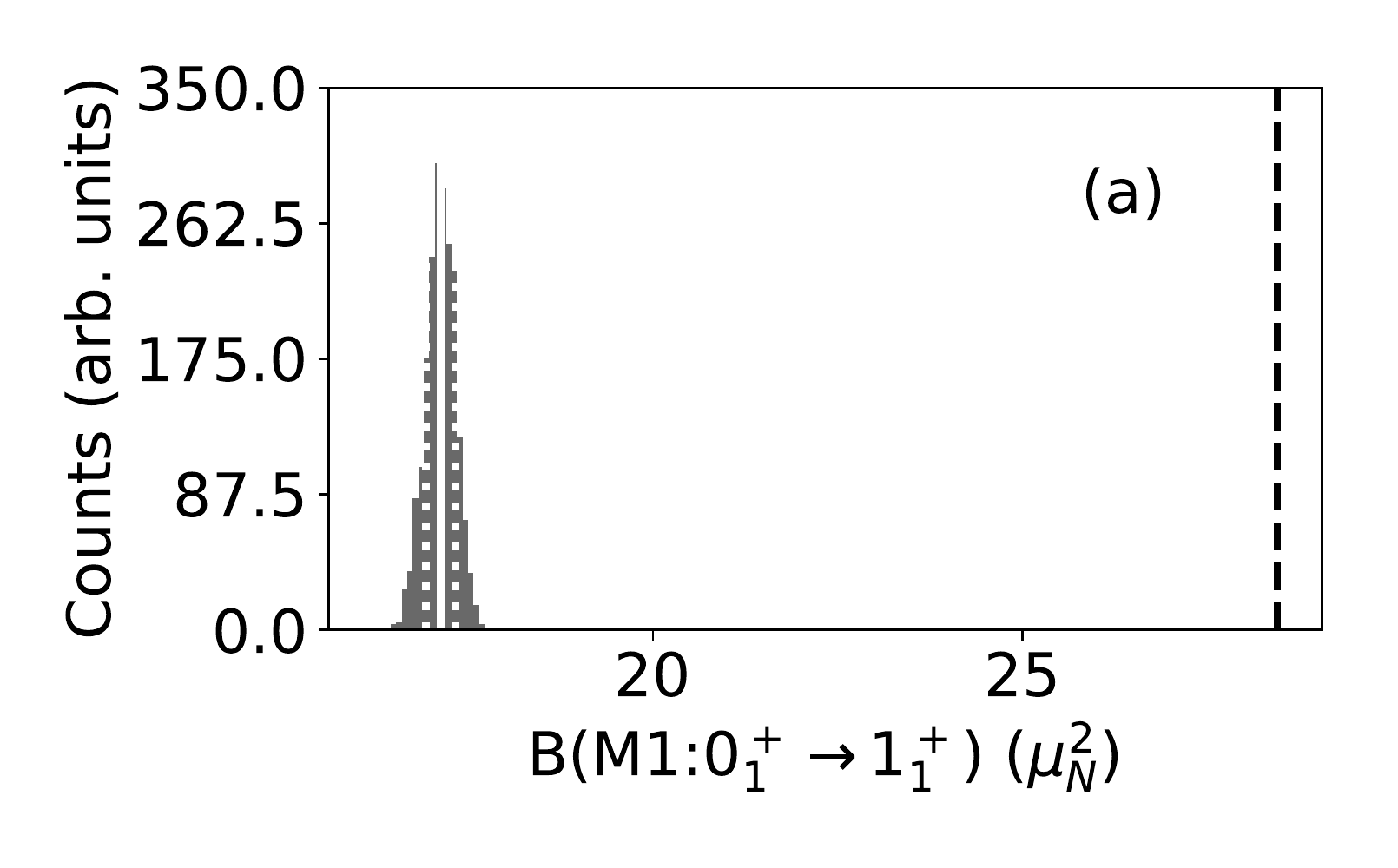} &   \includegraphics[scale=0.5,clip]{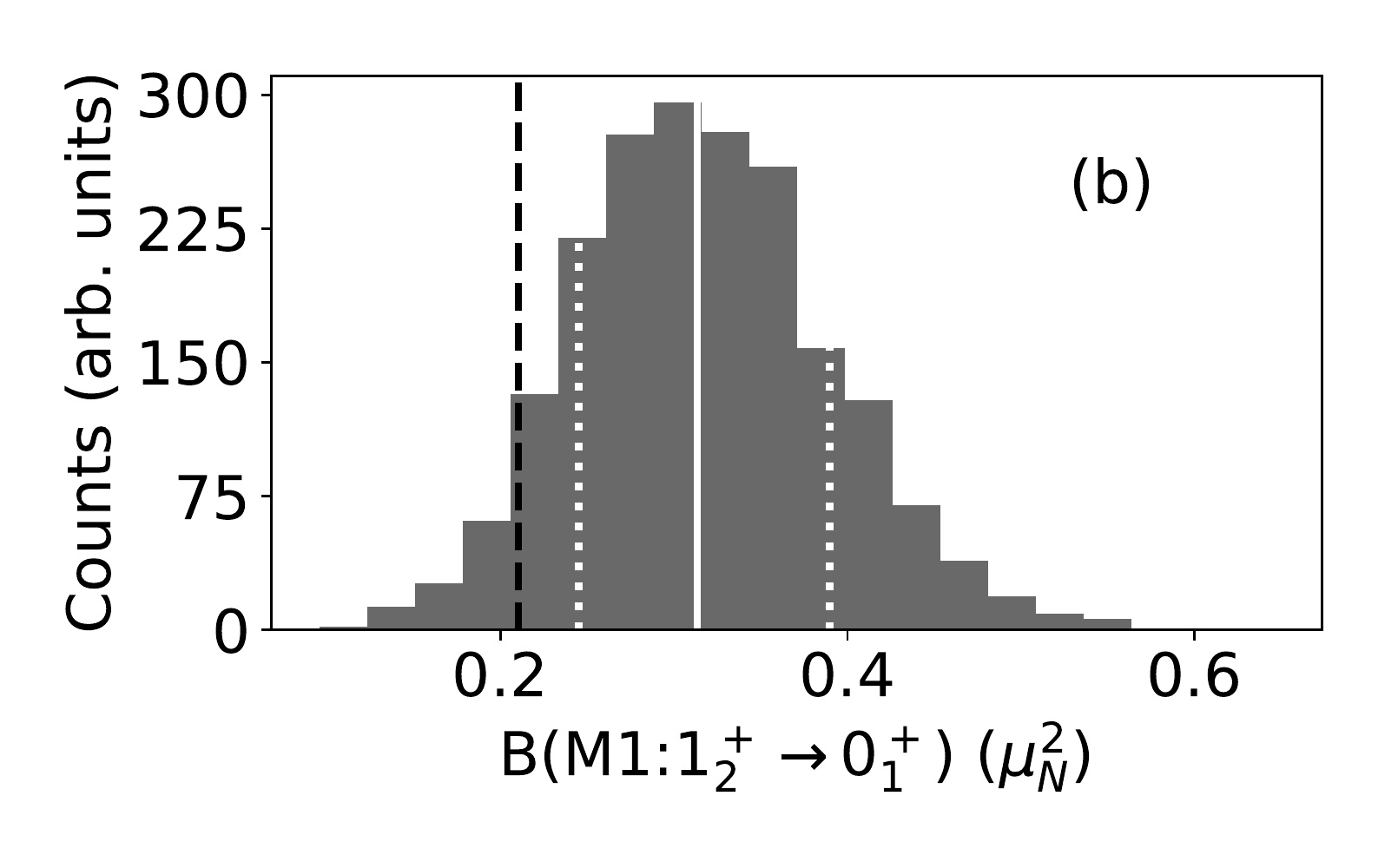} \\
     \end{tabular}
    \includegraphics[scale=0.5,clip]{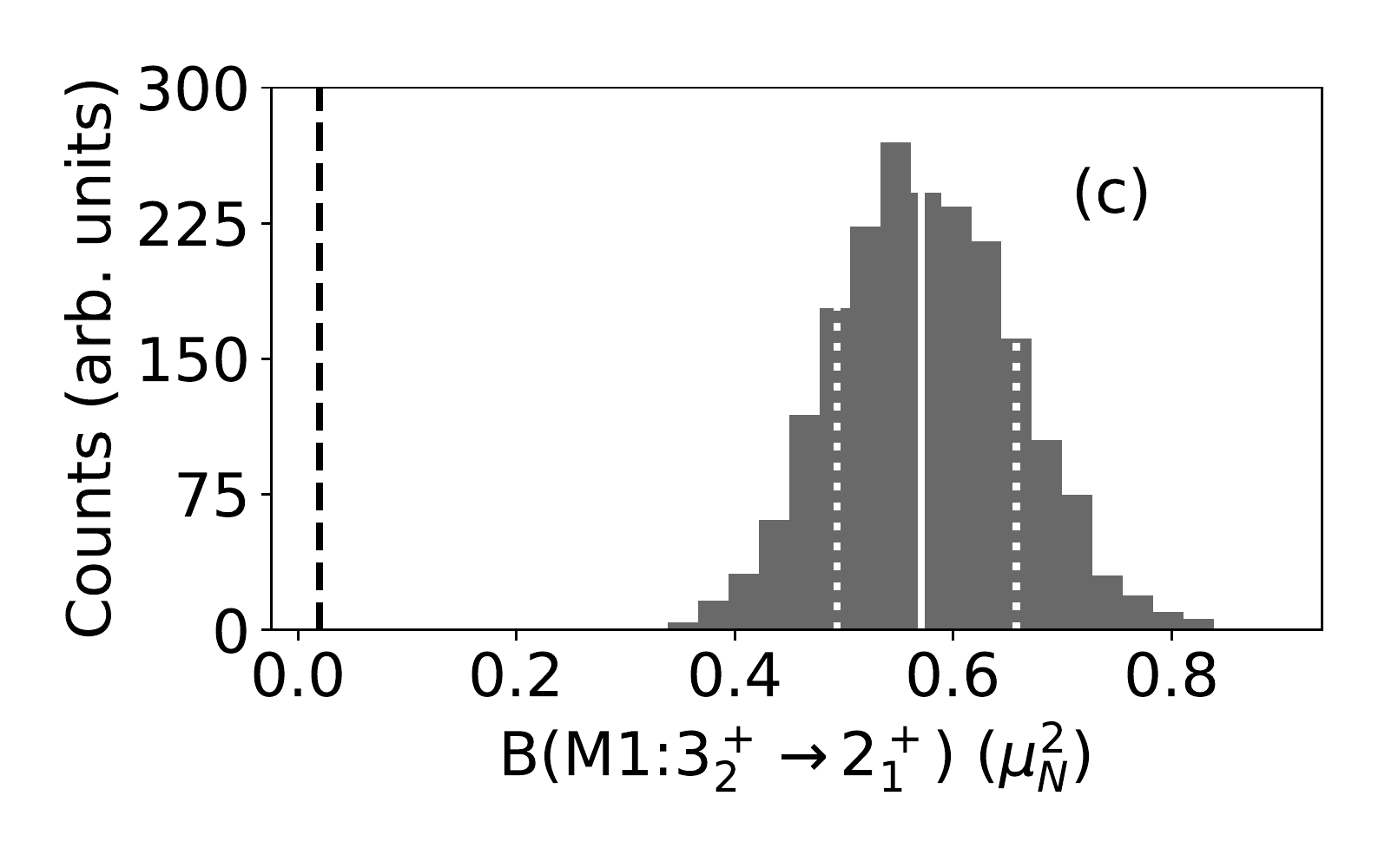}    \\
    \caption{Distributions of the magnetic dipole transition strengths for $^{18}$F. Black dashed line shows experimental value \cite{TILLEY19951}.  The uncertainty interval is highlighted in white: (a) $0_1^+ \rightarrow 1_1^+$ : $17.13 ^{ +0.19 }_{ -0.21 }$, (b) $1_2^+ \rightarrow 0_1^+$ : $0.31 ^{ +0.076 }_{ -0.068 }$, and (c) $3_2^+ \rightarrow 2_1^+$ : $0.57 ^{ +0.087 }_{ -0.077 }$, all in units $\mu_N^2$.}


    
    \label{fig:M1F18}
\end{figure}
\begin{figure}
    \begin{tabular}{cc}
      \includegraphics[scale=0.5,clip]{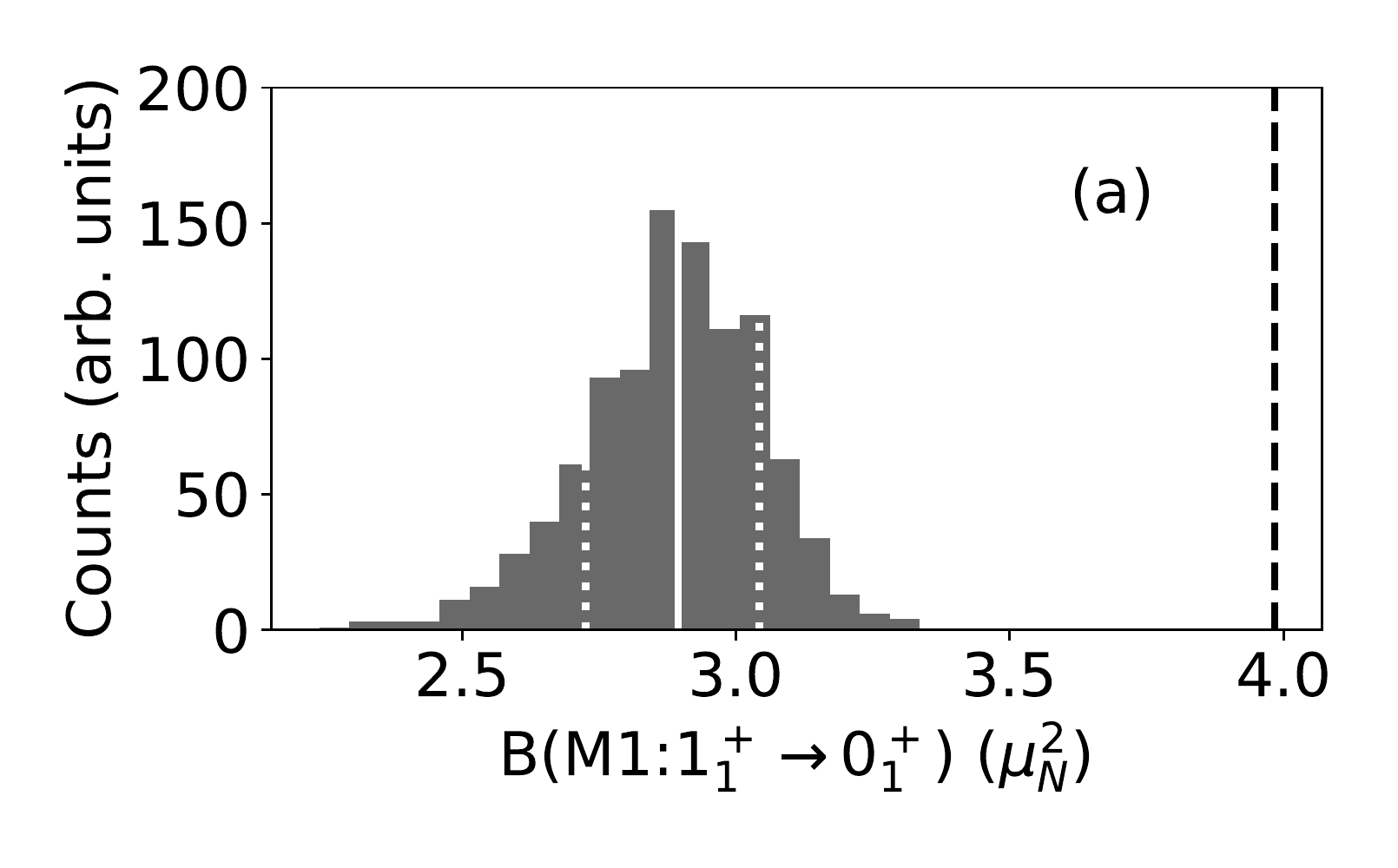} &  
      \includegraphics[scale=0.5,clip]{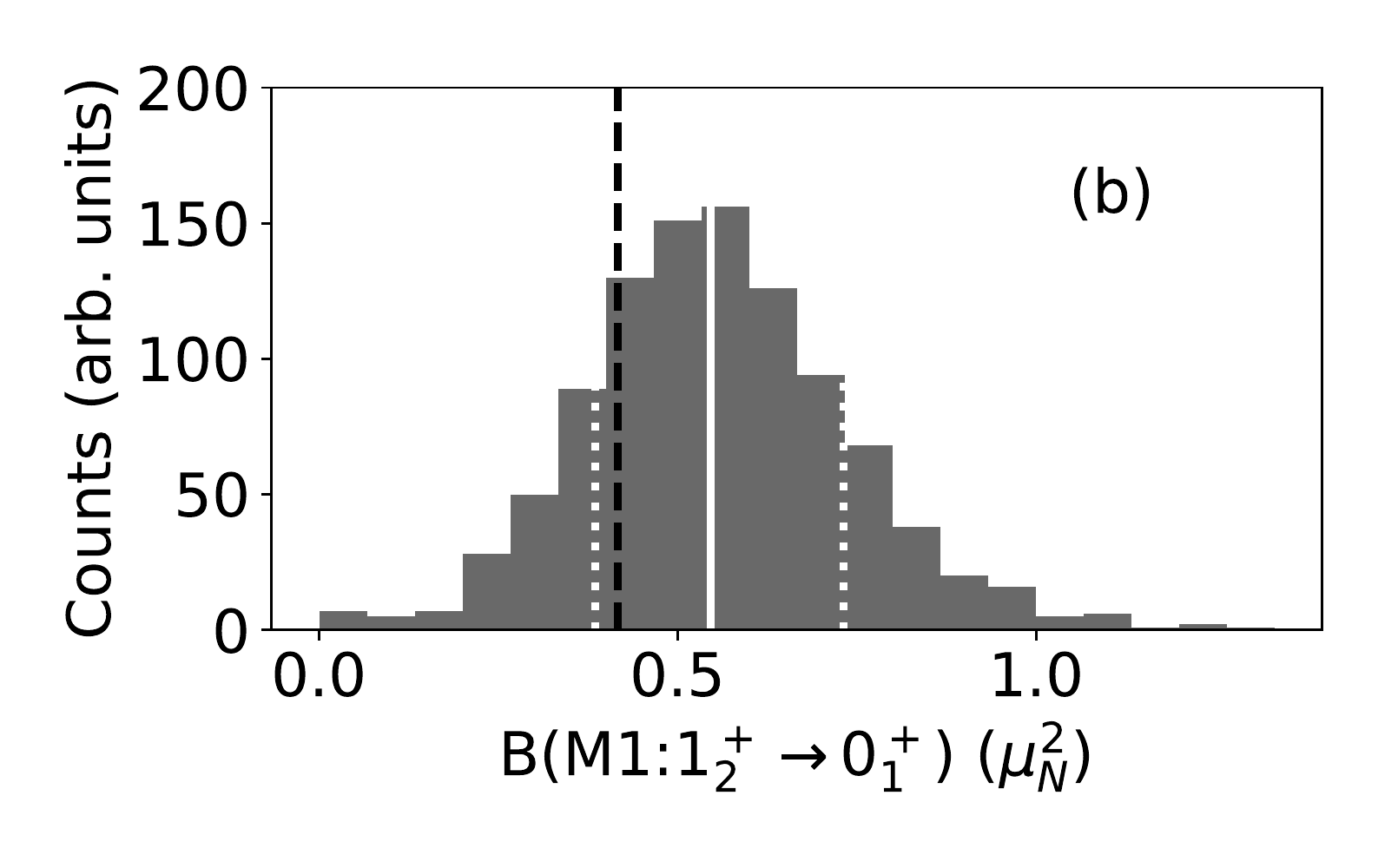} \\ \includegraphics[scale=0.5,clip]{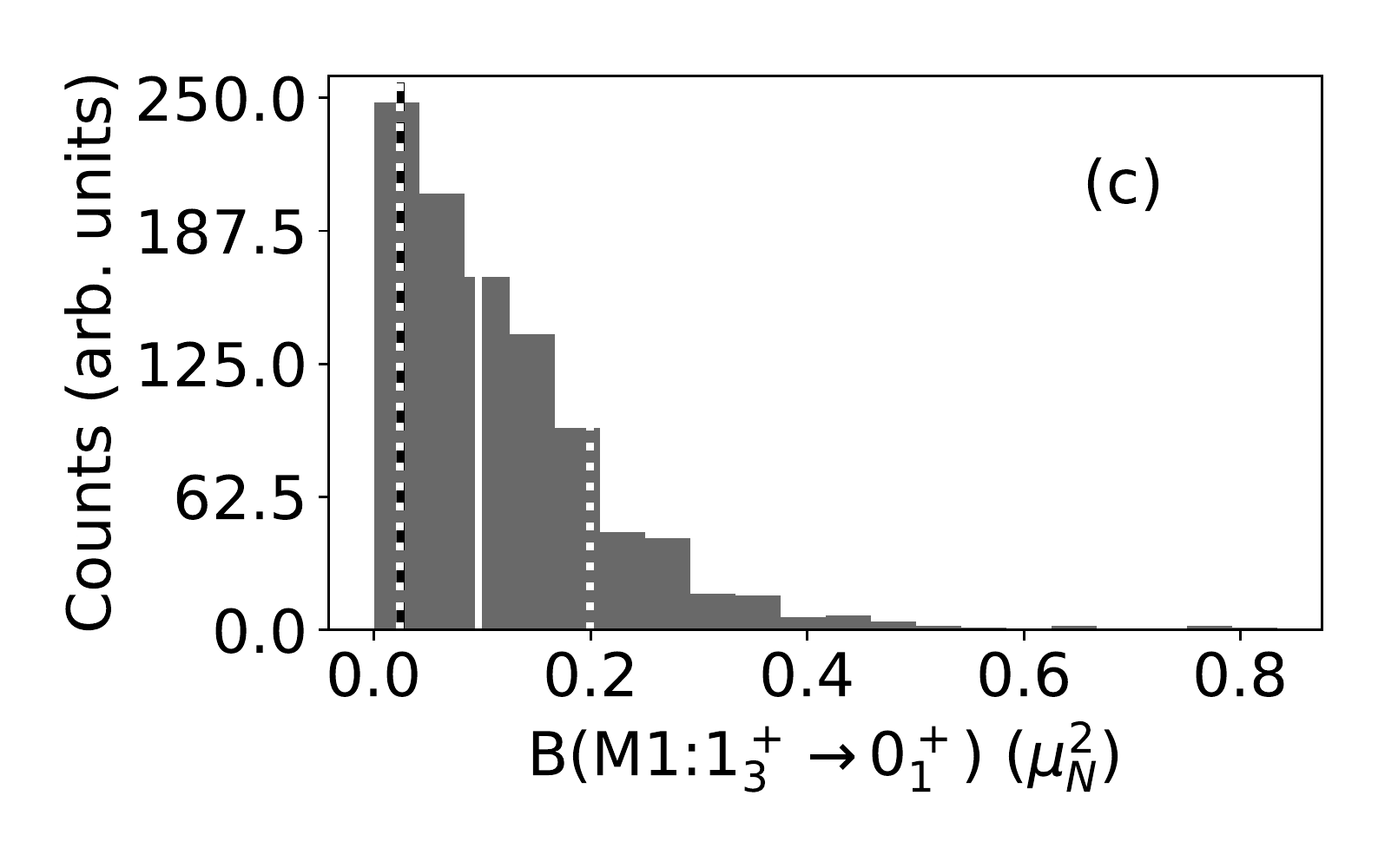} &  
      \includegraphics[scale=0.5,clip]{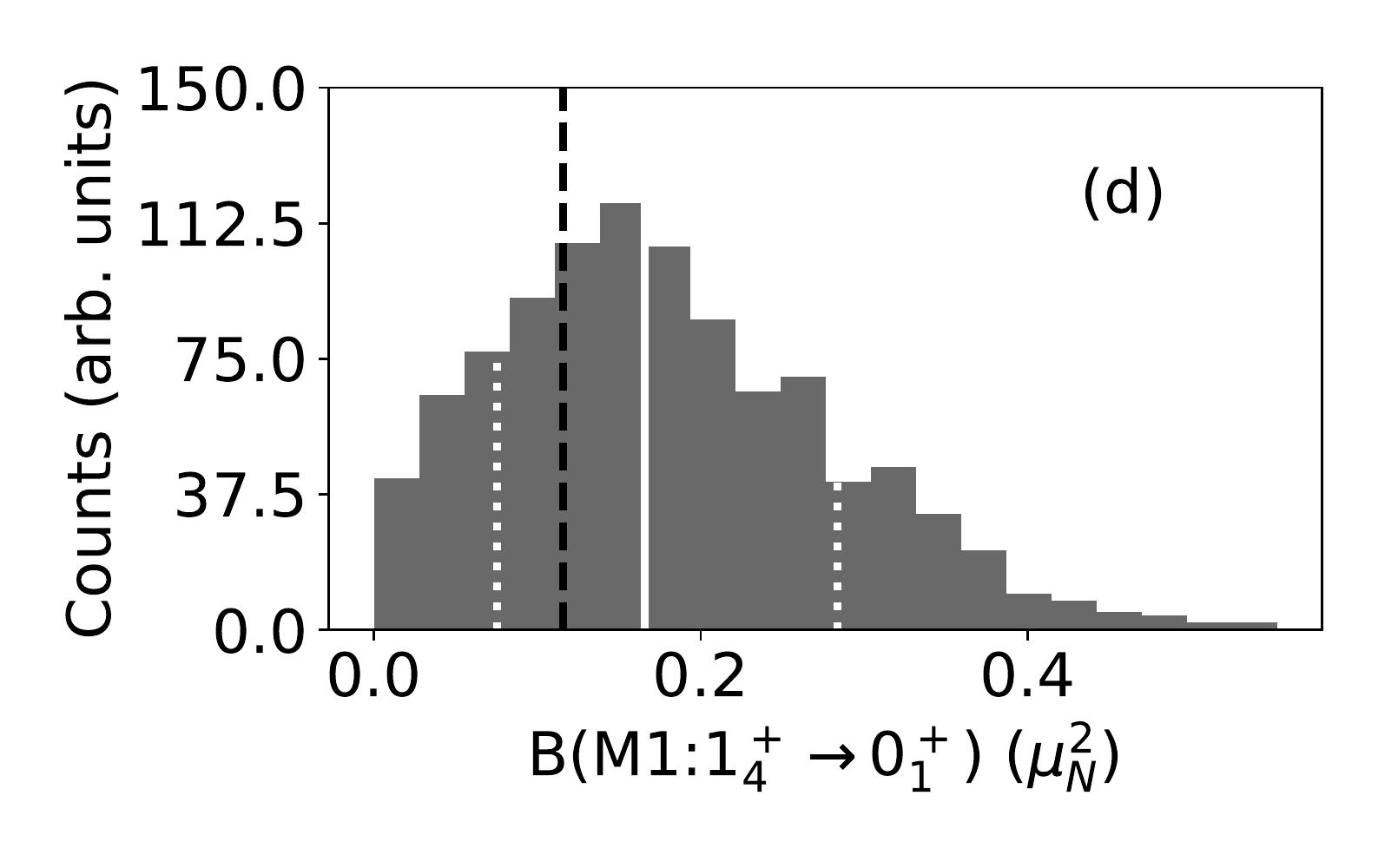} \\  
     \end{tabular}
    \includegraphics[scale=0.5,clip]{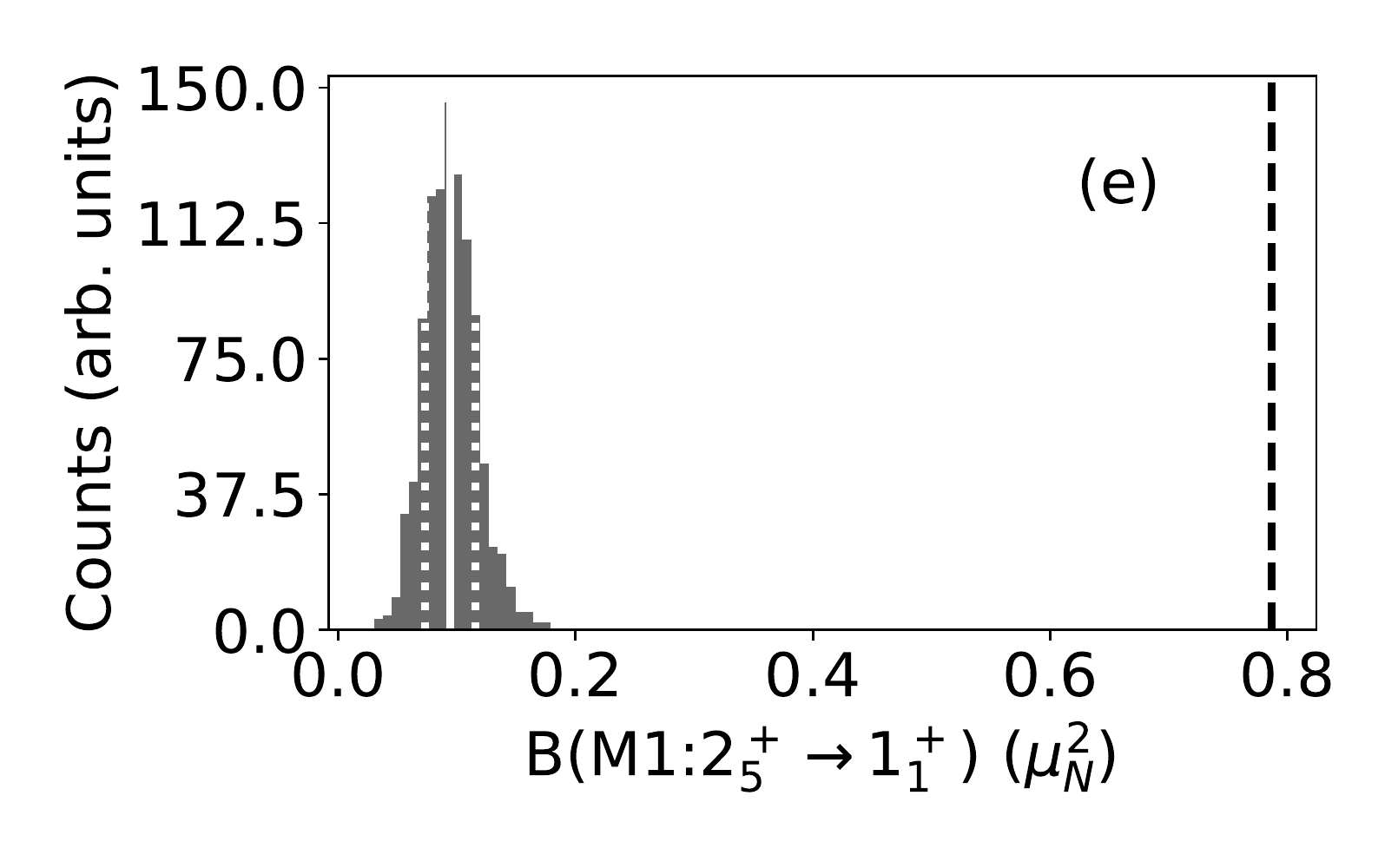}    \\
    \caption{Distributions of the magnetic dipole transition strengths for $^{26}$Al. Black dashed line shows experimental value \cite{BASUNIA20161}.  The uncertainty interval is highlighted in white: (a) $1_1^+ \rightarrow 0_1^+$ : $2.89 ^{ +0.15 }_{ -0.17 }$, (b) $1_2^+ \rightarrow 0_1^+$ : $0.55 ^{ +0.18 }_{ -0.16 }$, (c) $1_3^+ \rightarrow 0_1^+$ : $0.096 ^{ +0.10 }_{ -0.07 }$, (d) $1_4^+ \rightarrow 0_1^+$ : $0.17 ^{ +0.12 }_{ -0.09 }$, and (e) $2_5^+ \rightarrow 1_1^+$ : $0.095 ^{ +0.022 }_{ -0.021 }$, all in units $\mu_N^2$.}




    
    \label{fig:M1Al26}
\end{figure}

Magnetic dipole reduced transition strengths B(M1) distributions for $^{18}$F and $^{26}$Al are shown in Fig. \ref{fig:M1F18} and \ref{fig:M1Al26} respectively. We used bare gyromagnetic factors,
with no corrections for exchange currents. Like 
the B(E2) values, some of the transitions are close to experiment, while 
the $0_1^+ \rightarrow 1_1^+$ in $^{18}$F is quite far away.
For $^{18}$F, in Fig.~\ref{fig:M1F18}, the median values and uncertainty intervals for our selected transitions are $0_1^+ \rightarrow 1_1^+$ : $17.13 \substack{ +0.19 \\ -0.21 }$, $1_2^+ \rightarrow 0_1^+$ : $0.31 \substack{ +0.076 \\ -0.068 }$, and $3_2^+ \rightarrow 2_1^+$ : $0.57 \substack{ +0.087 \\ -0.077 }$, all in units $\mu_N^2$, where $\mu_N$ is the nuclear magneton, while for $^{26}$Al, in Fig.~\ref{fig:M1Al26} the median values and uncertainty intervals for our selected transitions are $1_1^+ \rightarrow 0_1^+$ : $2.89 \substack{ +0.15 \\ -0.17 }$, $1_2^+ \rightarrow 0_1^+$ : $0.55 \substack{ +0.18 \\ -0.16 }$, $1_3^+ \rightarrow 0_1^+$ : $0.096 \substack{ +0.10 \\ -0.07 }$, $1_4^+ \rightarrow 0_1^+$ : $0.17 \substack{ +0.12 \\ -0.09 }$, and $2_5^+ \rightarrow 1_1^+$ : $0.095 \substack{ +0.022 \\ -0.021 }$ .

\begin{figure}
    \begin{tabular}{cc}
      \includegraphics[scale=0.5,clip]{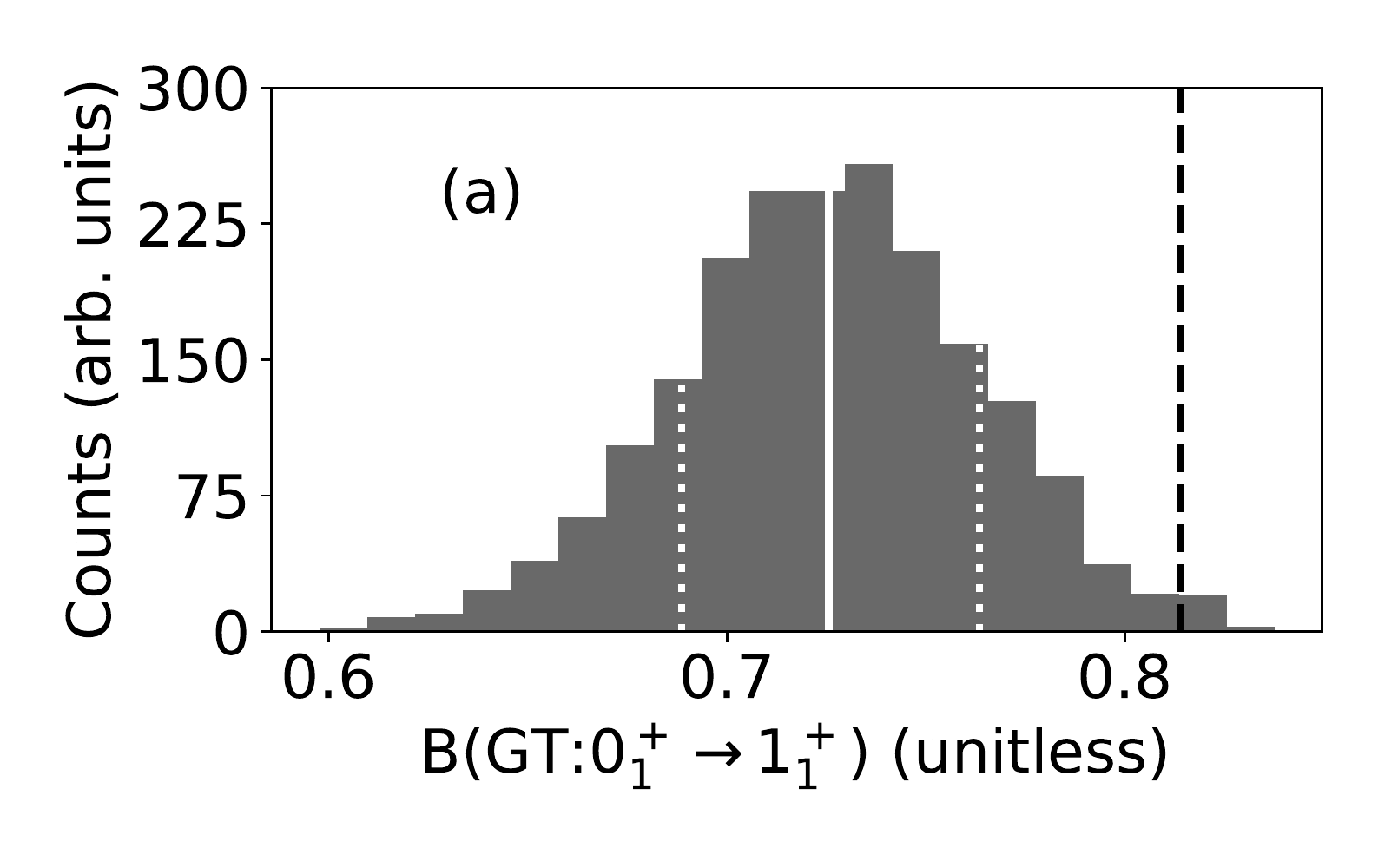} &   \includegraphics[scale=0.5,clip]{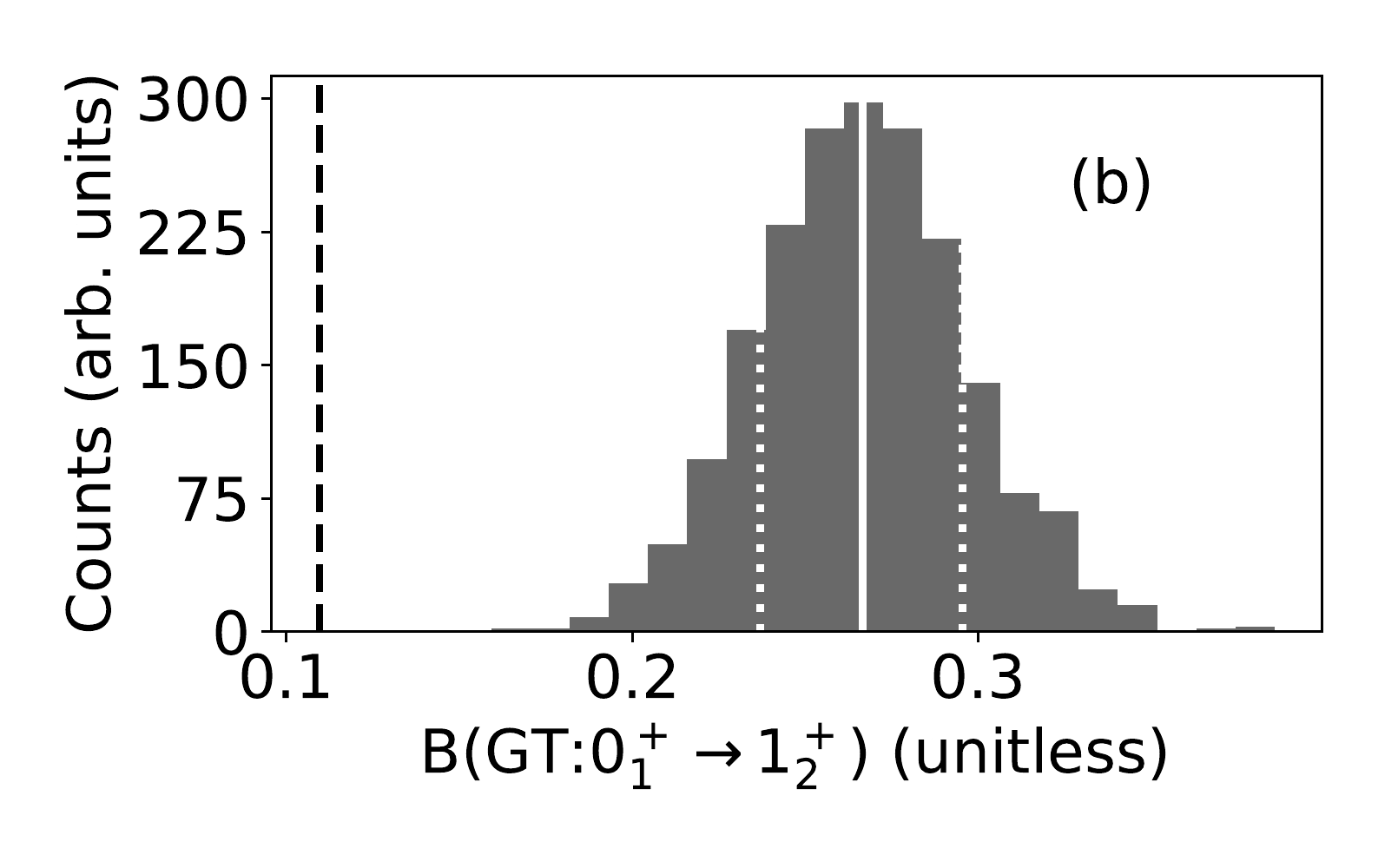} \\
    \end{tabular}
    \includegraphics[scale=0.5,clip]{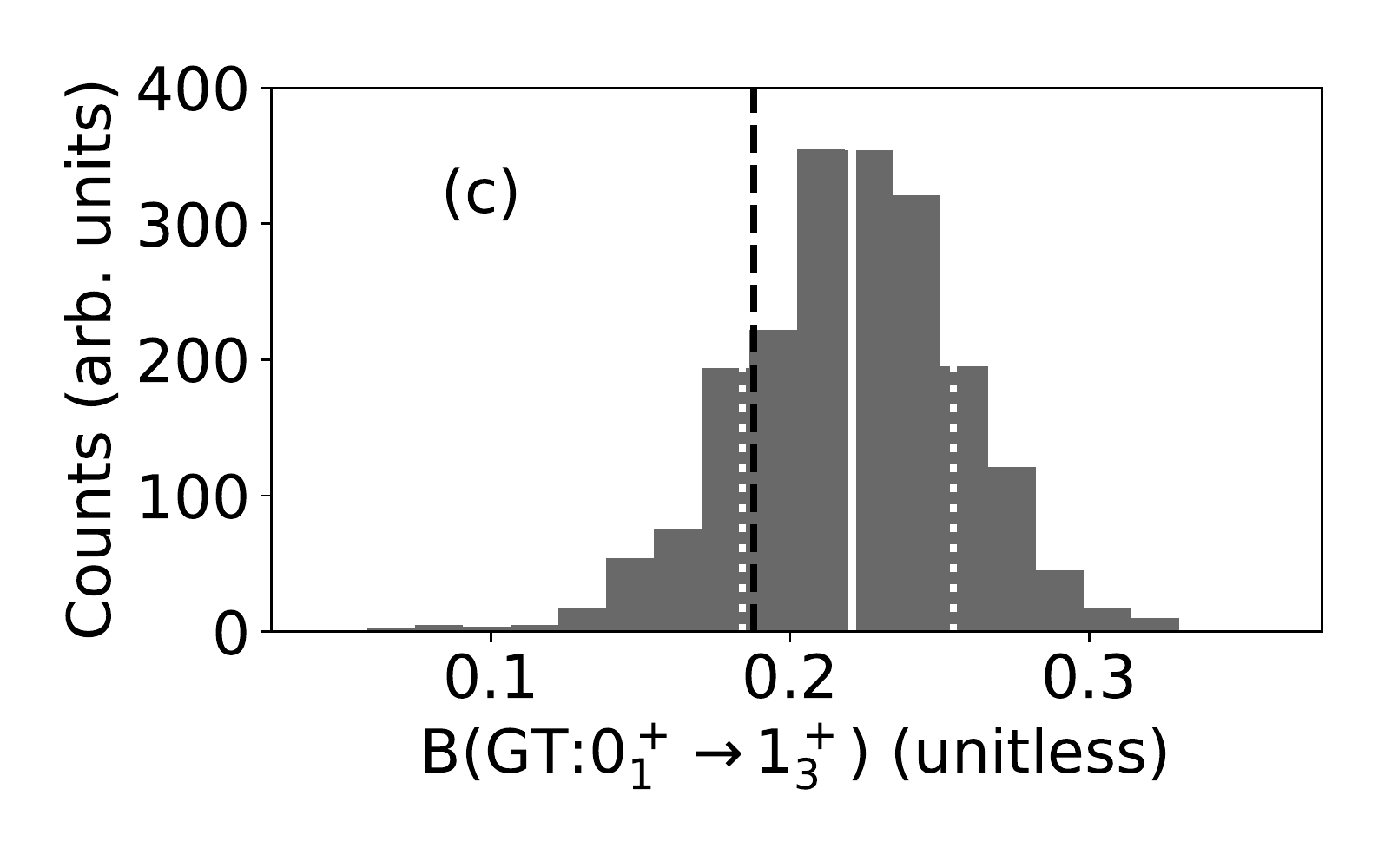}    \\
    \caption{Distributions of the Gamow-Teller (GT) transition strengths for $\beta^-$-decay of $^{26}$Ne to $^{26}$Na. Black dashed line shows experimental value \cite{BASUNIA20161}. The uncertainty interval is highlighted in white: (a) $0_1^+ \rightarrow 1_1^+$ : $0.726 ^{ +0.038 }_{ -0.037 }$, (b) $0_1^+ \rightarrow 1_2^+$ : $0.267 ^{ +0.029 }_{ -0.030 }$ ,and (c) $0_1^+ \rightarrow 1_3^+$ : $0.22 ^{ +0.034 }_{ -0.037 }$.}



    \label{fig:GTNe26}
\end{figure}

\begin{figure}
    \begin{tabular}{cc}
      \includegraphics[scale=0.5,clip]{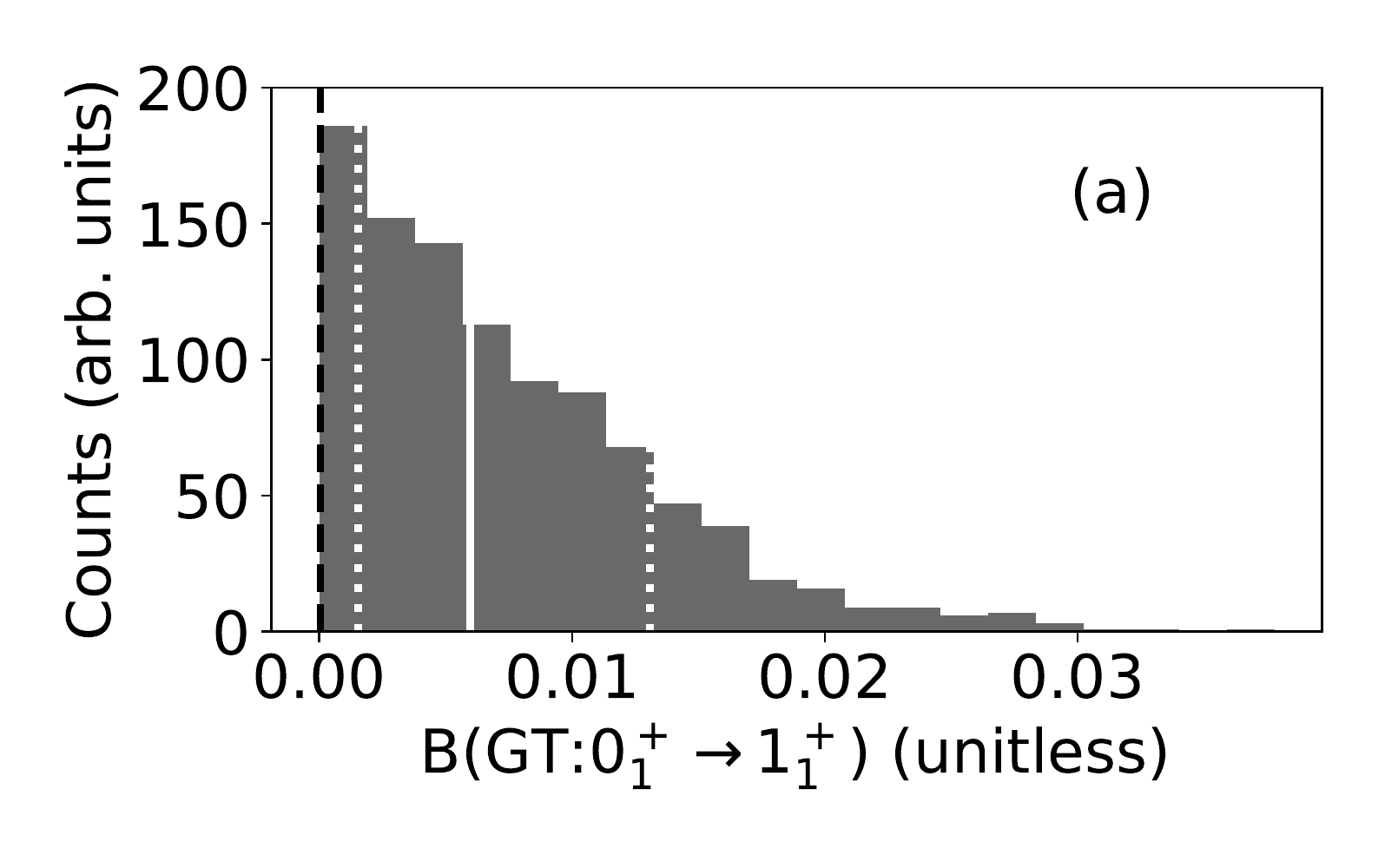} &   \includegraphics[scale=0.5,clip]{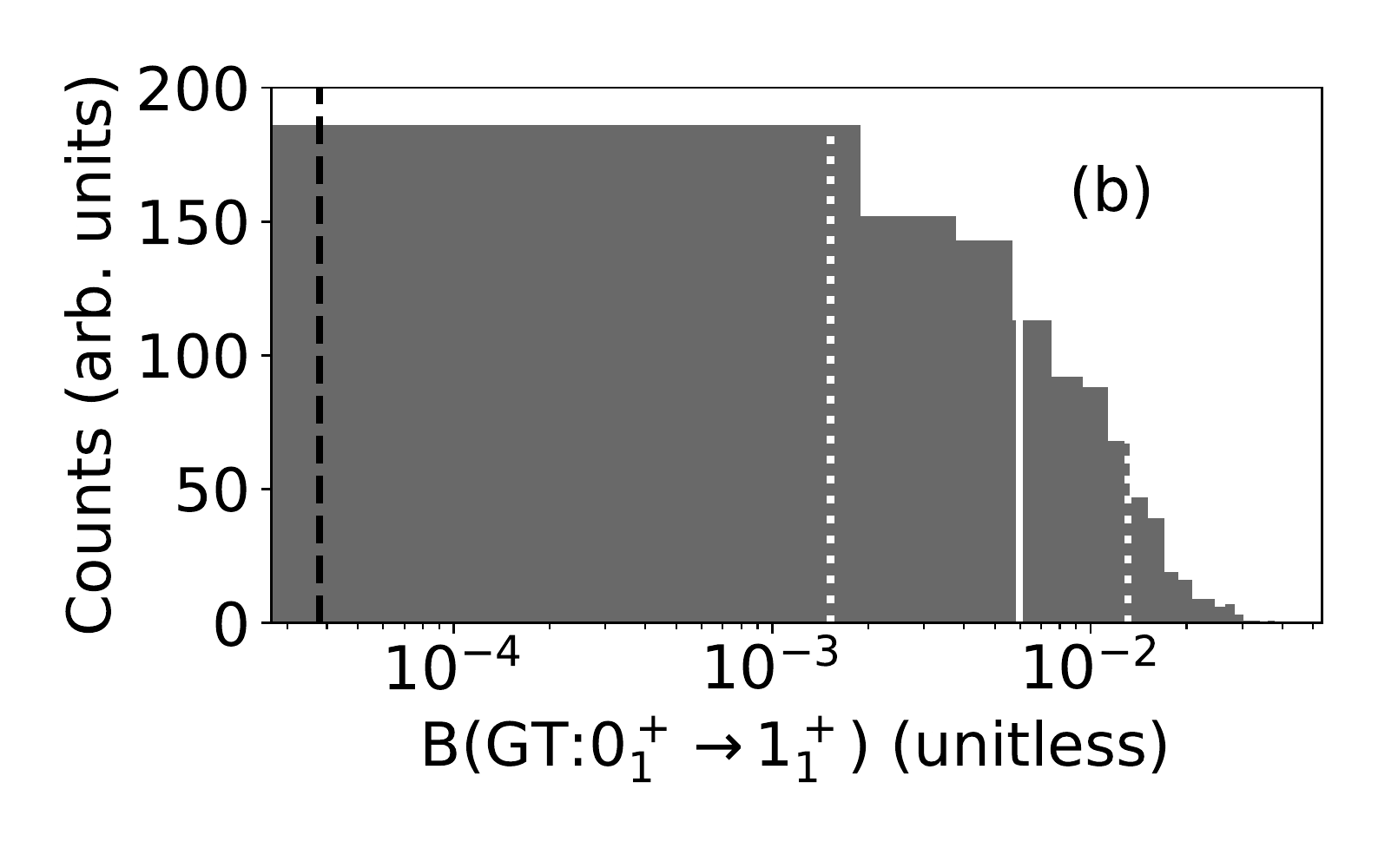} \\
    \end{tabular}
    \caption{Distribution of the Gamow-Teller (GT) transition strength for $\beta^-$-decay of $^{32}$Si to $^{32}$P ($0_1^+ \rightarrow 1_1^+$). The left plot is a linear scale in B(GT) and the right is log-scale. Black dashed line shows experimental value of $0.000038$ \cite{OUELLET20112199}. The uncertainty interval is highlighted in white: $0.00597 ^{ +0.0071 }_{ -0.0045 }$.\\  }
    
    \label{fig:GTSi32}
\end{figure}

We show Gamow-Teller matrix elements 
for $\beta^-$-decays in $^{26}$Ne and $^{32}$Si in Fig. \ref{fig:GTNe26} and \ref{fig:GTSi32} respectively. We have used for the axial-vector coupling constant $g_A / g_V = -1.251$, following \cite{PhysRevC.78.064302}, and a quenching factor of $0.76$ for USDB. 
For $^{26}$Ne, in Fig.~\ref{fig:GTNe26}, the median values and uncertainty intervals for our selected transitions are $0_1^+ \rightarrow 1_1^+$ : $0.726 \substack{ +0.038 \\ -0.037 }$, $0_1^+ \rightarrow 1_2^+$ : $0.267 \substack{ +0.029 \\ -0.030 }$ ,and $0_1^+ \rightarrow 1_3^+$ : $0.22 \substack{ +0.034 \\ -0.037 }$, all unitless. 
The ground-state decay of $^{32}$Si has a small experimental transition strength, so our sensitivity analysis does not provide a normal distribution for B(GT). Using USDB, our median value and uncertainties are $0.00597 \substack{ +0.0071 \\ -0.0045 }$, but this is quite different than the experimental value is of $0.000038$ \cite{OUELLET20112199}.
This particular transition is very sensitive to the parameters: 
for the 1985 universal $sd$-shell interaction (USD) interaction \cite{brown_sd_GT_1985} we get a value for B(GT) = $0.00005$, and 
if one uses the 2006 universal $sd$-shell interaction version A (USDA), 
which is a less constrained version of USDB  \cite{PhysRevC.74.034315},
the B(GT) is $0.038$.  

(Motivated by the non-Gaussian distribution in Fig.~\ref{fig:GTSi32}, we increased 
the number of samples from 1000 to 4000. The results were nearly indistinguishable,
with new median value and uncertainties of  $0.00624 \substack{ +0.0077 \\ -0.0047 }$.)

\bigskip

One of the biggest questions in physics today is the nature of non-baryonic dark matter
\cite{RevModPhys.90.045002}. While there are a number of ongoing and planned 
experiments \cite{roszkowski2018wimp}, interpreting experiments, including limits, requires 
good knowledge of the dark matter-nucleus scattering cross-section, including uncertainties. While historically 
it was assumed dark matter would couple either to the nucleon density or spin density, 
more recent work based upon effective field theory showed there should be a large number of 
low-energy couplings, around 15 \cite{PhysRevC.89.065501}.   This enlarged landscape of couplings, and the increased 
need for good theory, is a strong motivation for the current work. 

\begin{figure}
    \includegraphics[scale=0.6,clip]{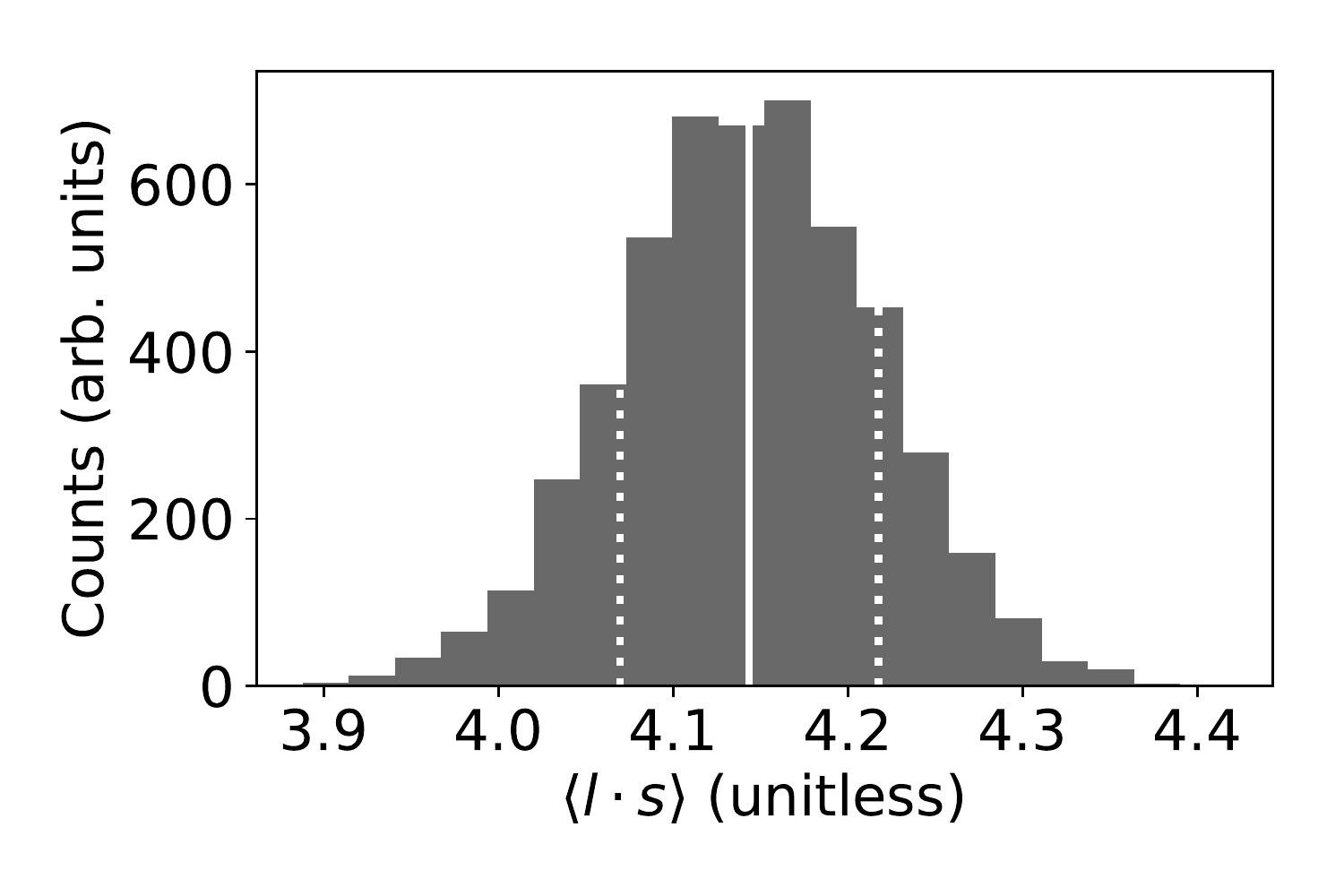} 
\caption{Distribution of $\langle l \cdot s \rangle$ in the ground-state of $^{36}$Ar. The $1\sigma$ interval is highlighted in white: $ 4.143 \pm 0.074$ .} 
\label{fig:ls}
\end{figure}

In order to illustrate the application of UQ to nuclear matrix elements for dark matter scattering,
Fig.~\ref{fig:ls} shows the uncertainty of an $\vec{l}\cdot \vec{s}$ coupling for $^{36}$Ar. 
$^{36}$Ar is a small component ($0.3\%$) of  argon dark matter detectors, e.g. \cite{agnes2015first}, but it is within the scope of the 
current work to compute.  Of the EFT operators that do not vanish for a $J^\pi = 0^+$ ground state,
most of them depend upon radial wave functions that do not play a role in fitting the USDB 
parameters; nontrivial operators, however, include $\vec{l}\cdot \vec{s}$,
which arises in
the long-wavelength (momentum transfer $q \rightarrow 0$) limit of the nuclear matrix elements of the operators ${\cal O}_{3,12,15}$  \cite{PhysRevC.89.065501}
\begin{eqnarray*}
{\cal O}_{3} = i \vec{S}_N \cdot \left ( \frac{\vec{q}}{m_N} \times \vec{v}^\perp \right ), \\
{\cal O}_{12}= \vec{S}_\chi \cdot \left ( \vec{S}_N \times \vec{v}^\perp \right ), \\
{\cal O}_{15}=- \left( \vec{S}_\chi \cdot \frac{\vec{q}}{m_N} \right ) 
\left (  \left ( \vec{S}_N  \times \vec{v}^\perp\right )    \cdot \frac{\vec{q}}{m_N}\right ),
\end{eqnarray*}
where $m_N$ is the nucleon mass, $\vec{q}$ is the momentum transfer, $\vec{S}_{N/\chi}$ are the 
spins of the nucleon/WIMP, and $\vec{v}^\perp$ is the component of the nucleon-WIMP relative 
velocity perpendicular to $\vec{q}$.
We chose to study $\langle \vec{l}\cdot \vec{s} \rangle$ 
for the simple reason of best illustrating a variance due to uncertainty in the USDB parameters. 
The variance of this particular operator is relatively small, but in larger model spaces 
there could be greater uncertainty.
Knowledge of the variance of the operator is important for interpreting experiments, such as 
placing upper limits on dark matter-nucleon couplings.

\section{Conclusions}

We have carried out uncertainty quantification of a `gold-standard' empirical interaction for
nuclear configuration-interaction calculations in the $sd$-shell valence.  Rather than finding the 
uncertainty in
each parameter independently \cite{PhysRevC.98.061301}, we computed the linear sensitivity of 
the energies, which is easy to compute using the Feynman-Hellmann theorem, and then constructed an approximate Hessian which we then diagonalized. This is 
equivalent to a singular-value decomposition of the linear sensitivity, and is also known as principle
component analysis. We found evidence  this is a good approximation to the full Hessian. 
From the inverse of the diagonal (in a basis of the PCA linear combination of parameters)
approximate Hessian, we obtained approximately independent 
uncertainties in the PCA parameters. 
Then, starting from those uncertainties, we generated uncertainties for energies as well as 
several observables.   The distribution of residuals in energies implies statistical agreement, as well as an 
underlying systematic uncertainty in the shell model of 150 keV.  For electromagnetic and weak transitions, 
which we note are sensitive to effective parameters such as effective charges and assumed oscillator length 
parameters, our residuals relative to experiment included both good agreement as well as residuals with statistically 
significant deviations.  We also presented as a test case a dark matter-nucleus interaction matrix element and our 
derived 
uncertainty.

In the Supplementary material\cite{supplementary}, we provide 
the list of energies, courtesy of B. A. Brown, to which USDB was fit, 
and
the eigenvalues and eigenvectors of our principal component analysis.

In future work, in addition to further and more systematic study of observables, we will carry out a more detailed and thorough study of parameter covariances, as well as applying our methods to other empirical interactions in other model spaces. This will entail, following Appendix A, evaluating the posterior without Laplace's approximation, and instead using Markov-chain Monte Carlo sampling. 
We are investigating the use of eigenvector continuation \cite{frame2018eigenvector,PhysRevLett.123.252501,konig2019eigenvector} to explore parameter space 
efficiently.
For the time being, however, it seems that this approximate Hessian is a good approximation. This is not surprising, but it is useful.  Nonetheless, moving to larger spaces, which grow exponentially in dimensions and compute time, will be challenging.  New technologies still in development, such as quantum computing may make possible
better and more rigorous uncertainty quantification.

\section{Acknowledgement}

We thank B.~Alex Brown for a helpful file of the energies to 
which USDB was fit, and S.~Yoshida for further discussion and encouragement to look at the full calculation of the 
Hessian. G.~W.~Misch suggested we look at the transitions of $^{26}$Al,
which have astrophysical relevance, and M.~Horoi pointed out
the $\beta$-decay of $^{32}$Si as a known case of sensitivity to the interaction.
This material is based upon work supported by the U.S. Department of Energy, Office of Science, Office of Nuclear Physics, 
under Award Numbers  DE-FG02-96ER40985 (for nuclear structure) and DE-SC0019465 (for dark matter 
matrix elements).  JF thanks the Computational Science Research Center for partial support.

\appendix

\section{The Bayesian context}


Our development above is cast in terms of standard 
sensitivity analysis.  To connect with more sophisticated UQ analyses, and 
to set the stage for future work, we provide a broader, Bayesian context.

To define uncertainty on the USDB parameters, we start with Bayes'
theorem. Let $D$ represent data and $\lambda$ the parameters, then
	\begin{equation}
		P(\bm{\lambda} | D) = \frac{P(D|\bm{\lambda})P(\bm{\lambda})}{P(D)} \propto P(D|\bm{\lambda}) P(\bm{\lambda})
	\end{equation}

Bayes' theorem states that the distribution of model parameters given the experimental data (the \textit{posterior} = $P(\bm{\lambda} | D)$) is proportional to the distribution of data (the \textit{likelihood} = $P(D|\bm{\lambda})$) given the parameter set, multiplied by the \textit{a priori} distribution of parameters (the \textit{prior} = $P(\bm{\lambda})$). 
Bayesian analysis \cite{sivia_bayes} demands that we put some thought into the choice of prior, and the typical choice here is a \textit{non-informative} prior, which seeks to minimize the effects of prior knowledge on the posterior distribution. In this case a non-informative prior can simply be uniform and very broad in the limiting case, $P(\bm{\lambda}) = \text{constant}$ everywhere. 
This assigns equal probability to all parameter values (the principle of indifference \cite{sivia_bayes}). 
Although one could also justify using an informative prior, the flat prior it is a sensible first approximation for the scope of this analysis.

With the prior set to constant, Bayes’ theorem reduces to:
\begin{equation}
        \label{reduced_bayes}
		P(\bm{\lambda} | D) \propto P(D | \bm{\lambda} )
\end{equation}

The goal now is to evaluate this expression, and we can choose between two methods: Laplace’s Approximation (LA), or Markov-Chain Monte Carlo (MCMC). Due to its simplicity, we choose LA, as did a 
prior shell model study \cite{PhysRevC.98.061301}. While MCMC advantageously makes no assumption as to the form of $P(\bm{\lambda}|D)$, it typically converges slowly for posteriors which are steep around extrema, so the computational cost of LA is comparatively much less. 


Laplace's approximation is a second-order Taylor approximation in the log-likelihood, and thus we  assume normally distributed errors on energies. Our likelihood function takes the form:
\begin{equation}
	P(D|\bm{\lambda}) = \exp \left[ - \frac{1}{2} \chi^2(\bm{\lambda}) \right]
	\label{likelihood}
\end{equation}
where $\chi^2$ is the usual sum of squared residuals: 
\begin{equation}
		\chi^2(\bm{\lambda}) = \sum_{\alpha=1}^{N} \left(  \frac{ E_\alpha^{SM}(\bm{\lambda}) - E_\alpha^{exp}}{\Delta E_\alpha} \right)^2
\end{equation}

$ E^{exp}_\alpha $ is the experimental excitation energy given in the data set and $ E^{SM}_\alpha(\bm{\lambda}) $ is the shell model prediction for that energy using the parameters $ \bm{\lambda} $, 
with 
 total uncertainty on the residual $\Delta E_\alpha $ (see discussion in Section \ref{sensitivity} and in particular
 Eq.~(\ref{add_errors})).

By Eq.~\ref{likelihood}, there exists a global maximum of this likelihood function, called the maximum likelihood estimator (MLE). The optimal point for the posterior is called the ``maximum \textit{a posteriori}'' (MAP), and here we see that $\bm{\lambda}_{MAP}=\bm{\lambda}_{MLE}$, but of course this is only in the special case of uniform prior. In this work, the MAP is equal to the USDB parameters.
\begin{equation}
	\bm{\lambda}_{\text{MAP}} = \argmax_{\bm{\lambda}} P(\bm{\lambda}|D) = \argmax_{\bm{\lambda}} P(D|\bm{\lambda})P(\bm{\lambda}) = \argmin_{\bm{\lambda}} \chi^2(\bm{\lambda}) = 
	\bm{\lambda}_\text{USDB}
\end{equation} 
The virtue of LA is we can immediately write down a properly normalized Gaussian approximation of the posterior: 
\begin{equation}
	P(\bm{\lambda} | D) \approx \frac{|H|^{1/2}}{(2\pi)^{k/2}} \exp \left[ - \frac{1}{2} (\bm{\lambda} - \bm{\lambda}_{\text{MAP}})^T H (\bm{\lambda} - \bm{\lambda}_{\text{MAP}}) \right],
\end{equation}
where $k$ is the dimension of the parameter space, and $H$ denotes the Hessian of the log-posterior (for brevity we refer to this as “the Hessian”). 
The Hessian is defined as minus the second-derivative (in $\bm{\lambda}$) of the log-likelihood about the MAP.

\begin{equation}
	H = - \nabla \nabla \log P(\bm{\lambda} |D) \rvert_{\bm{\lambda} = \bm{\lambda}_{\text{MAP}}}
\end{equation}

Because of Eq.~\ref{reduced_bayes}, we can introduce an arbitrary constant $ c $, so       $ P(\bm{\lambda}|D) = cP(D|\bm{\lambda}) $:
\begin{equation}
H = - \nabla \nabla \log P(\bm{\lambda} |D)  = - \nabla \nabla \log c P(D|\bm{\lambda}) = 0 - \nabla \nabla \log P(D|\bm{\lambda}) = + \frac{1}{2} \nabla \nabla \chi^2 (\bm{\lambda}) ,
\end{equation}
so the elements of $ H $ become:
\begin{equation}
	H_{ij} =  \frac{1}{2} \frac{\partial^2 \chi^2 (\bm{\lambda})}{\partial \lambda_i \partial \lambda_j}
\end{equation}

Under these assumptions, we proceed as described in the main text.




\section{Computed covariance of fitted energies}

Here we show that computing the covariance matrix of fit energies $C_E$ by Eq. \ref{energy_cov} is simply related to a similarity transform of the original 
uncertainties on  fit energies given by Eq. \ref{add_errors}: $\Sigma_{\alpha \alpha} = \Delta E_\alpha $ .
The response of the energies to changes in the parameters is an $N_d \times N_p$ Jacobian matrix, $J_{\alpha i} = \partial E_\alpha / \partial \lambda_i$,  where $N_d$ is the number of data points and $N_p$ is the number of parameters.
The approximate Hessian is 
\begin{equation}
    A = J^T \Sigma^{-2} J    ,
\end{equation}
and the parameter covariance is 
\begin{equation}
    C_\lambda = A^{-1} = (J^T \Sigma^{-2} J)^{-1}.
\end{equation}
Since $J$ is not square, we cannot evaluate this expression in terms of matrix inversion and instead use the pseudoinverse obtained by SVD decomposition. We get the factorization $J = USV^T $ where $U$ is a $N_d \times N_d$ unitary matrix, $S$ is a $N_d \times N_p$ matrix with the only non-zero elements being $N_p$ singular values along the diagonal, and $V$ is a $N_p \times N_p$ unitary matrix. We use this to define a new matrix $J^+$ which is the pseudoinverse of $J$.
\begin{equation}
    J^+ = VS^+U^T 
\end{equation}
Here, $S^+$ is the pseudoiverse of $S$, which has the same shape as $S^T$ and the only nonzero elements are such that $S^+_{jj} = 1/S_{jj}$ for $j = 1,2,..., N_p$.

Plugging this into the expression for $C_\lambda$ we have
\begin{equation}
    C_\lambda = J^+ \Sigma^2 [J^T]^+ = (V S^+ U^T) \Sigma^2 (U S^+ V^T),  
\end{equation}
In turn we insert this into our expression for $C_E$:
\begin{equation}
    C_E = (U S V^T) (V S^+ U^T) \Sigma^2 (U S^+ V^T) (V S U^T)  .
\end{equation}
By the orthogonality of $U$ and $V$ we have $U^T U = I_{d}$ and $U^T U = I_{p}$, identity matrices in the data-space and parameter-space respectively, so that
\begin{equation}
    C_E = U S I_{p} S^+ U^T \Sigma^2 U S^+ I_{p} S U^T  .
\end{equation}
To simplify further, we need to pay attention the the rank-deficient property of $S$. Define $S I_{p} S^+ = P_{d}^{p}$ to be a $N_d \times N_d$ square matrix with $N_p$ 1's on the diagonal, starting from the top, and all zeros otherwise. (This is projection operator from the data-space into the parameter-space, hence this notation.) 
Then
\begin{equation}
    C_E = U P_{d}^{p} U^T \Sigma^2 U P_{d}^{p} U^T  .
\end{equation}
Now, notice that since $\Sigma^2$ is diagonal, we have $U^T \Sigma^2 U = \Sigma^2$. The matrix $P_{d}^{p}$ is of course idempotent so $ P_{d}^{p} P_{d}^{p} = P_{d}^{p} $, and we get
\begin{equation}
    C_E = U \Sigma^2 P_{d}^{p} U^T  , 
\end{equation}
or
\begin{equation}
    U^T C_E U = \Sigma^2 P_{d}^{p}   .
\end{equation}
Thus, the computed covariance on the energies $C_E$ is equivalent to a similarity transform of the input 
uncertainties $\Sigma^2$, albeit with  rank $ = N_p$.


\section{The rotated quantile-quantile plot}
The quantile-quantile (Q-Q) plot \cite{nist_stats_book} is a useful tool for visualizing how well the distribution of a data set matches that of a random variate from a known probability distribution. Our \textit{rotated} Q-Q plot in Fig. \ref{fig:qqplot} shows the comparison of energy residuals to a standard normal distribution. The following gives a brief explanation.

A typical Q-Q plot graphs $N$ measured data points $\{ x^{\mathrm{data}}_i \}$, sorted from lowest to highest, against $N$ uniformly distributed evaluations $\{ x^\mathrm{eval}_i \}$ of the \textit{quantile function} (sometimes called a \textit{percent-point function}) of the distribution we wish to compare to. For a random variable $X$ with cumulative distribution function (CDF) $F_X(x) \equiv \text{Pr}(X \le x)$, the quantile function $Q_X(p)$ returns the value of $x$ such that $F_X(x)=p$; in other words, it is the inverse function of the CDF. For instance if the set of data points follows a normal distribution, that is, 
$\{ x^\mathrm{eval}_i \} = \{ x^\mathrm{normal}_i \}$ 
then the points $(x_i^{\text{data}},x_i^{\text{normal}})$ for $i=1,2,\dots,N$ will fall on a straight line with slope of 1. If the data does not follow a normal distribution, then the points will deviate from a straight line, 
displaying how non-normal the data is. Our Q-Q plot in Fig. \ref{fig:qqplot} in this paper has been ``rotated'' by plotting instead
$(x_i^{\text{data}}-x_i^{\text{normal} },x_i^{\text{normal}})$, where 
$x^{\text{data}}$ are the energy residuals, so that  a normal 
distribution would lie on the horizontal axis at zero. This allows for an easier identification of discrepancies between empirical and theoretical quantiles via visual inspection. \par
Many statistical tests exist for determining normality of data, and often these can be represented as a curve on the Q-Q plot. The Kolmogorov-Smirnov and tail-sensitive tests used in this work correspond to curves shown in Fig. \ref{fig:qqplot}; evidence of possible non-normality of the data is indicated by the plotted quantile-quantile points crossing over these curves.

\bibliography{johnsonmaster,foxmaster,navarroperez,exp_data}

\end{document}